# Skeletonization of neuronal processes using Discrete Morse techniques from computational topology

Samik Banerjee[1], Caleb Stam[2], Daniel J. Tward[3], Steven Savoia[1], Yusu Wang[2], Partha P.P.Mitra[1*]

[1]Cold Spring Harbor Laboratory, Cold Spring Harbor, 11724, New York, USA.
[2]University of California San Diego, La Jolla, 92093, California, USA.
[3]University of California Los Angeles, Los Angeles, 90095, California, USA.

*Corresponding author(s). E-mail(s): parthaxmitra@gmail.com;

**Abstract**

To understand biological intelligence we need to map neuronal networks in vertebrate brains. Mapping mesoscale neural circuitry is done using injections of tracers that label groups of neurons whose axons project to different brain regions. Since many neurons are labeled, it is difficult to follow individual axons. Previous approaches have instead quantified the regional projections using the total label intensity within a region. However, such a quantification is not biologically meaningful. We propose a new approach better connected to the underlying neurons by skeletonizing labeled axon fragments and then estimating a volumetric length density. Our approach uses a combination of deep nets and the Discrete Morse (DM) technique from computational topology. This technique takes into account nonlocal connectivity information and therefore provides noise-robustness. We demonstrate the utility and scalability of the approach on whole-brain tracer injected data. We also define and illustrate an information theoretic measure that quantifies the additional information obtained, compared to the skeletonized tracer injection fragments, when individual axon morphologies are available. Our approach is the first application of the DM technique to computational neuroanatomy. It can help bridge between single-axon skeletons and tracer injections, two important data types in mapping neural networks in vertebrates.

## 1 Summary

Neuroscientific data analysis has traditionally involved methods for statistical signal and image processing, drawing on linear algebra and stochastic process theory. However, digitized neuroanatomical datasets containing labeled neurons, either individually or in groups labeled by tracer injections, do not fit into this classical framework. The tree-like shapes of neurons cannot be adequately described as points in a vector space (e.g., the subtraction of two neuronal shapes is not a meaningful operation). There is therefore a need for new approaches, which has become more pressing given the growth in whole-brain datasets in which axons and dendrites are labeled via sparsely labeled neurons or tracer injections.

Methods from computational topology and geometry are naturally suited to the analysis of neuronal shapes. In this paper, we introduce methods from Discrete Morse Theory to skeletonize neuronal processes or process fragments from tracer-injected brain image data, leading to a summarization of the neuronal projections based on the volumetric line density of such skeletonized processes in space. This contrasts with previous approaches in which the neuronal projections are quantified by counting fluorescently labeled voxels. Such a procedure is difficult to connect to the underlying biology, except in a qualitative manner. In contrast, our skeletonization process allows us to carry our a biologically more meaningful quantification, in terms of the length-density of the neuronal process fragments in given regions of space. The total length of axons is biologically more meaningful than the label density as it can provide information about the number of presynaptic sites on the axons in different brain compartments[1]. The total length of all axons emanating from the injection site in a given brain compartment or across the whole brain can be obtained by interpolating and integrating the volumetric line density sampled in a series of optical planes, as is normally the case for whole-brain light microscopic imaging of tracer injections. This also provides a way to relate the tracer injection data quantitatively to single axon reconstructions, which are increasingly available in some vertebrate animals, particularly in the laboratory mouse.

The proposed algorithmic procedure for neuron skeletonization includes an initial process detection step [2], which applied to a projection region of a tracer-injected brain image volume produces a likelihood map of the neural processes or process fagments. This is effectively a normalization and preprocessing step. This first step is followed by extraction of the process skeletons from the density field using the Discrete Morse technique, in which the 1-unstable manifold of the likelihood function, which connects the local maxima through intervening saddle points, is extracted using persistent homology as a noise-control method. This structure, which is part of the Morse skeleton of the likelihood function, could be intuitively interpreted as tracing a path through a mountainous landscape by connecting the tops of adjacent hills connected by tall ridges.

After suitable corrections to the raw skeletons designed to take into account the underlying biological structure of the data, we find that this procedure leads to an effective skeletonization of neuronal process fragments, in tracer injected whole brain microscopic image data, as long as they are not too dense. This then permits the direct quantification of the lengths of the process fragments which is suitable for further summarization of the tracer injections. We apply the method to high-resolution brain image data of tracer-injection labeled neurons collected using two different microscopic techniques, demonstrating better performance than a baseline algorithm using non-topological methods (significant improvement in precision, and significant speedup in proofreading).

The extracted process skeletons are then passed to a summarization stage, where the length density of the fragments is quantified in the two-dimensional image plane. This provides a bridge between single neuron skeletons and tracer injection data. Single neuron skeletons are increasingly available, mapped to a common atlas coordinate space. Such single neuron skeletons could be virtually sliced into thin sections corresponding to the optical sections obtained in the microscopic imaging of tracer injection data, and the length density of the fragments thus obtained can be quantified in the two-dimensional plane of section. This gives rise to a length density as a function of space, which could directly be compared with the length density obtained from the tracer-injection labeled fragments. We illustrate this procedure through an example. Further, availability of the voxelixed line-density for the tracer injection data, allows us to compute an information theoretic measure of how much extra information is provided by single neuron reconstructions over tracer injections in a brain region, which should help clarify the relation between the tracer injection labeled groups of neurons and the constitutent individual neurons.

The DM method can trace a neuronal process fragment through regions of low intensity as it utilizes the global topological structure present in the data. Persistent homology based simplification of the Morse skeleton allows the method to deal with noise in the data in an adaptive manner, by considering local differences in intensities rather than absolute intensity values. Additionally, the DM approach is theoretically principled and conceptually clean, minimizing multiple ad-hoc hand-engineered steps. On the other hand, there is a significant computational overhead to the topological data analysis approach, however we are able to mitigate the speed issues by using parallelized implementations of the Discrete Morse algorithm.

## 2 Background

**Topological Data Analysis and Discrete Morse Theory:** Topological data analysis (TDA) methods have been applied across domains to analyze complex high-dimensional datasets [3]. The benefit of TDA methods is to study the structure that depends on the connectivity properties of the data independent of specific metrical and geometrical properties. TDA use a variety of approaches to characterize the topological structure underlying the data in question[4–8]. Some of the relevant computational tools (in particular persistent homology [9]) have been applied to multiple subject domains [10–14], including neuroscience [15–17].

The subarea of TDA pertinent to the present work is a persistence-guided Discrete Morse theory-based computational framework for reconstructing hidden graphs from observed data. Discrete Morse theory has been utilized to capture hidden structure in 2D or 3D volumetric data [18–20]. The extraction of hidden graphs was formulated in [21], and the framework was simplified and theoretical guarantees provided in [22]. Morse theory based methods are sensitive to the global topology of the data in contrast with methods sensitive only to local structure. Thus, for example, the underlying graph skeleton of a noisy measurement of a scalar field can be traced through regions of weak signal. Morse theory in its original form applies to continuous functions on manifolds. Discrete Morse theory [23] is a discretized and combinatorial computational framework inspired by Morse theory, suitable for algorithmic implementation on digitized data. Persistent homology is used to separate signal from noise and remove potentially noise-related structure from the graph. The pertinent TDA tools have previously been applied to the reconstruction of hidden road networks from noisy GPS trajectories and satellite images [24, 25]. Here we adapt and extend this approach to develop a computational methodology suitable for computational neuroanatomy and to address neuroscientific problems.

In this manuscript we introduce a data analysis framework entitled DM-skeleton that uses TDA and the Discrete Morse approach to skeletonize groups of neurons labelled by tracer injections. For tracer injection skeletonization, DM-skeleton provides a conceptually new route to the analysis and quantification of mesoscale projection data, and shows robust performance.

**Tracer Injection Skeletonization:** In tracer injected brain image volumes, thousands of neurons with somata or terminals co-localized in a brain compartment are collectively labeled using the tracer injection, and the individual neurons cannot generally be skeletonized. Conventionally, brain-wide connectivity information is summarized in the form of regional connectivity matrices [26]. Such a representation loses connection with the axonal morphology of individual neurons and is difficult to interpret in more microscopic terms. Here we introduce a new approach to the analysis of tracer injection data, by *skeletonizing* axon fragments labeled by the tracer injections using discrete morse method. This permits us to quantify the local length density of the labeled axons, which can then be further related to the length density of underlying single axons. To the best of our knowledge, this is a new approach to conceptualizing tracer-injection data, and could provide a biologically better-grounded approach to the study of mesoscale connectivity mapping using tracer injections. One important advantage of the tree-skeletonization approach is that one can then estimate the total length of all the neurons labeled by the tracer injection. While we do not expect that estimates based on sampled fragments to be as precise as one would get by actually tracing whole axons, single axon tracings are expensive and time intensive to obtain, and simulations have shown that there is reasonable correlation between projected length fragments with actual axon lengths[1]. Therefore we believe that a careful skeletonization of axon fragments from tracer injections and subsequent estimation of length densities is methodology worth developing.

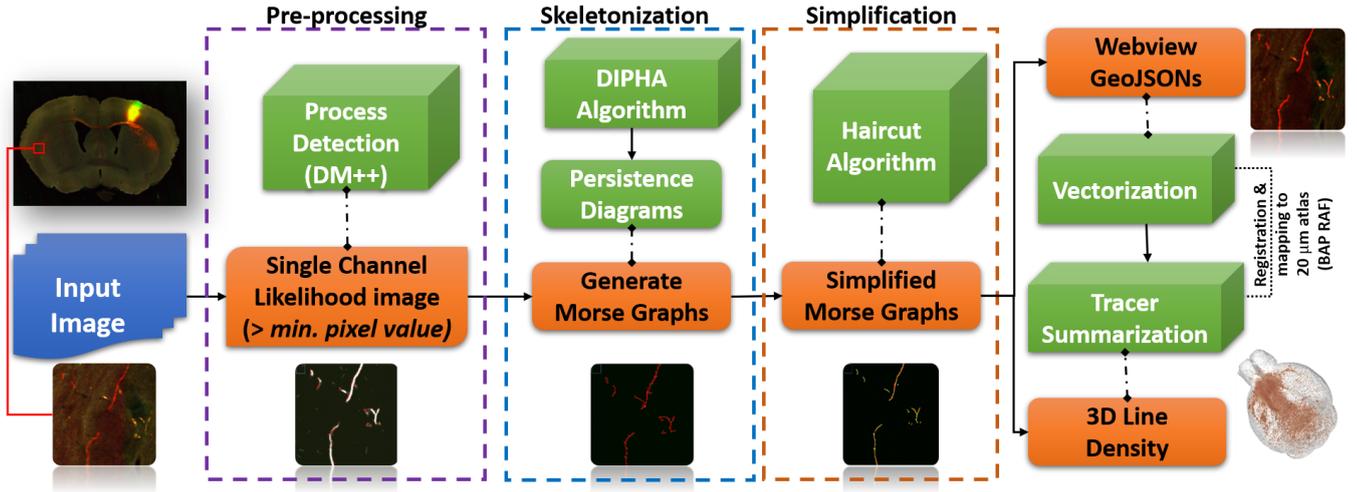

**Fig. 1** / Workflow and sample outputs for the DM-Skeleton pipeline. Workflow for 2D Skeletonization (from sparse neuronal label to individual skeleton extraction) and Summarization of the Detected Labels (from dense neuronal label to injection summarization) with sample outputs for each step.

The DM pipeline for summarizing multiple-neuron tracer injection datasets has multiple steps. First, the raw image stack is preprocessed to detect neuronal processes as a likelihood map, which we do using a previously introduced method combining deep networks and topological data analysis[2]. Then a variant of the Discrete Morse algorithm[22, 24] is used to produce a graph skeleton containing all potential axon fragments. As a noise reduction step, a persistent homology based simplification step is carried out next. The denoised graph is next further processed to extract a minimal spanning tree taking into account the biological prior knowledge that axons have tree-like topology. We provide the resulting pipeline as a computational package that takes images as inputs and produces a DM-skeleton data structure consisting of the detected axon fragments as output (see https://data.brainarchitectureproject.org/pages/skeletonization for code and data).

## 3 Results

**Method overview.** The workflow of the proposed DM-skeleton (DM-Skeleton) method is shown in Fig. 1. The workflow has three main steps, namely preprocessing, skeletonization, and simplification (see the Methods section 1.3.)

DM-Skeleton takes 2D scalar images as input. In Step 1, the DM++ algorithm[2] is applied as a normalization step, generating a likelihood image. The likelihood image, which is interpreted as a normalized label density field $\rho$, is used as a Morse function which serves as an input to the next stage. The goal of the next step is to capture centerlines passing through (relatively) high likelihood regions using the 1-unstable manifold of the density function (see Fig. 2 for an explanatory graphic).

In Step 2, a persistence-guided discrete Morse-based framework [22, 24] is applied to $\rho$, producing as an output the Morse 'graph skeleton' $G$. The 1-unstable manifold connects peaks through saddles, thus bridging through low-density regions along the labeled axon fragments (e.g., gaps and weak signals along the Y-junction in Fig. 2a). Finally, in Step 3, the Morse graph skeleton $G$ is further processed to extract axon fragments. First, false positives are suppressed by intersecting with a binary mask created by appropriately thresholding the likelihood to remove very low probability regions. The maximal spanning trees of the resulting graph fragments are then extracted. As noted above, the tree fragments occasionally have easily identified spurious side-branches ("hair"). We remove such short side branches using a simple *"Haircut"* algorithm; more details in Methods section 1.3.3).

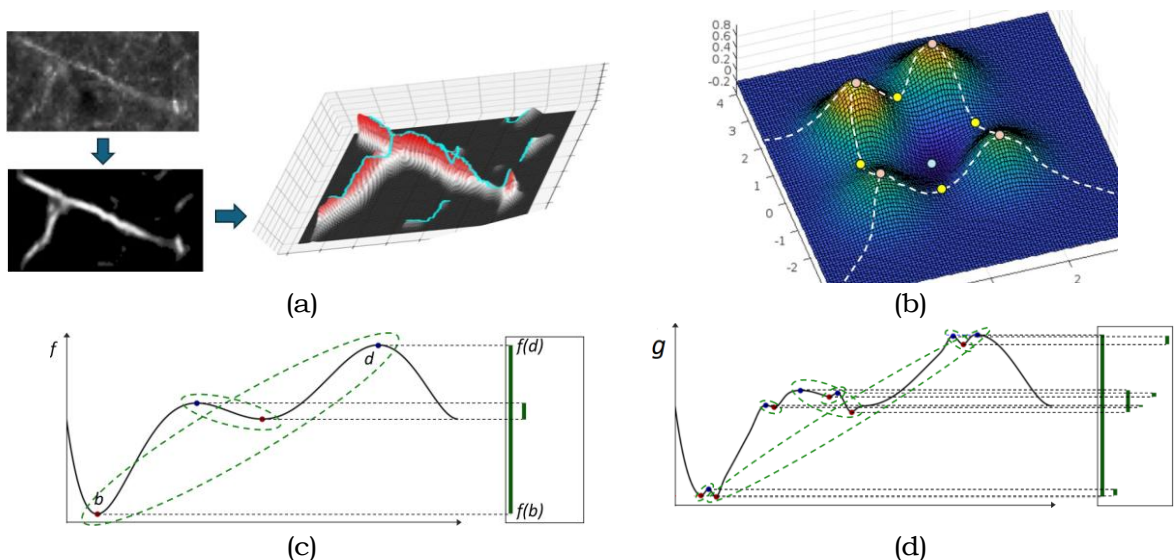

**Fig. 2 /** Illustration and basic concepts used for Discrete Morse theory based graph skeletonization algorithm. **(a)**. An input raw image (top left) is first converted to a likelihood image (left bottom). Treating the likelihood map as a density function (the corresponding terrain is show on the right), extract the 1-unstable manifold of this function, which is a one dimensional branched structure that traces paths following the gradients in the density, connecting peaks through intervening saddle points. **(b)** An example of A 2D Morse function together with the Morse skeleton (white dashed curves): pink points are local maxima, yellow points are saddles, while blue points are local minima. The Morse skeleton is the collection of the so-called 1-unstable manifolds (integral paths of gradient descent dynamics connecting saddles to maxima / mountain peaks). **(c)** Persistence is used to remove small or noise peaks as a denoising step. An example of persistence pairs on a simple 1D function $f: \mathbb{R} \to \mathbb{R}$, represented by the so-called *persistent barcodes* given by the lifetimes of features as the function value is smoothly increased, shown as the vertical green segments on the right. In particular, given $f$, as we gradually increase $f$-function values, topological features (in this case connected components) first appear (are 'born') at local minima, and disappear ('die') at local maxima. Each persistence pair (*i.e.* $(b, d)$) indicates the birth and death of some feature (*i.e.* born at $f(b)$ and killed at $f(d)$). This gives rise to a interval $(f(b), f(d))$ (shown as vertical bars) forming the so-called persistence barcode w.r.t.$f$ (on the right). The 'persistence' of the feature $(b, d)$ is defined to be the difference in function values $|f(d) - f(b)|$ which can be considered as a measure of stability, i.e. how "long" the feature persists / lives. In the persistence barcode, the persistence of a feature $(b, d)$ is defined to be the length of the corresponding persistent bar. In the function plots, persistence pairings are marked by green dotted curves. The function $g$ in **(d)** can be viewed as a noisy perturbation of function $f$. The function $f$ has 2 prominent features (persistence pairs), while the perturbed version $g$ also has additional "smaller" features with lower persistence (corresponding to very short persistent bars in the right).

To illustrate the proposed technique on real data and to compare with a baseline algorithm, we processed digitized microscopic images of two brain volumes, corresponding to two tracer-injected brains with tracer injections placed at nearby locations, imaged with two different microscopy methods, namely Whole Slide Microscopy using fluorescent imaging (WSI) [27], and Serial Two Photon microscopy (STP) [28–30]. As a baseline algorithm we used the function *bwskel()* from MATLAB that uses the medial axis transform, which is a standard approach to skeletonization of similar shapes. This function attempts to provide a topology-preserving thinning of the object o be skeletonized, and therefore provides a reasonable baseline comparison of our topologically motivated Morse-theory based skeletonization method. Likelihood images corresponding to the detection of axon fragments in these brain images, obtained using the *DM++* technique which we have previously described[2], were binarized using *OTSU*-based threshold. This binarized likelihood image was skeletonized using the *bwskel()* function from the MATLAB Image Processing Toolbox. Comparisons of the DM-skel output and baseline bwskel output on the same input images can be seen in Fig.3 and Fig.4 respectively for the WSI and STP imaged brains.

Fig.3 shows images from the WSI data set whereas Fig.4 shows images from the STP data set. In both figures, columns (a) and (c) show tiles showing tracer-labeled axon fragments from the WSI data set, with selected zoomed-in regions shown in columns (b) and (d). The top row shows the original fluorescent imaging data, the second row the results of the proposed method (DM-skel), and the third row the results of the baseline method (bwskel).

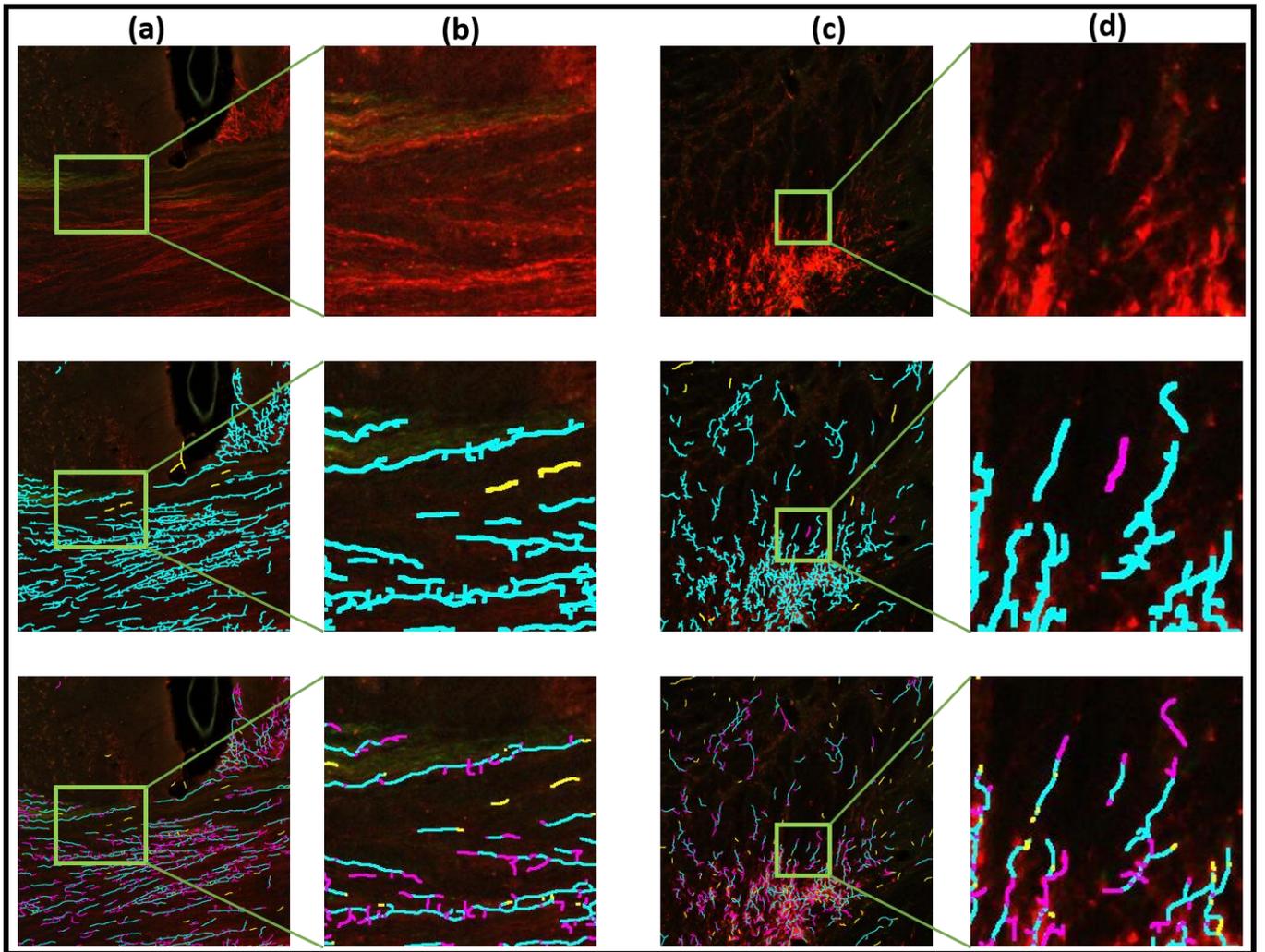

**Fig. 3 / Axon skeletonization for fluorescent Whole Slide Image (WSI) data.** Columns (a),(c) show image tiles and columns (b),(d) show zoom ins. Top row shows the original images, the middle row the results of DM-skeleton, and the bottom row the baseline skeletonization results using the MATLAB function bswkel. Axon fragments are seen in the original image as groups of lines of enhanced intensity. Results of visual-manual annotations for errors are also shown. *True-positives* are in *cyan*, *false-positives* are in *yellow* and *false-negatives* are in *magenta*. The results show better continuity and detection of the fragments by the DM-skeleton method over the baseline method.

From visual inspection, we can see that the baseline method bwskel (shown in the third rows of the respective figures) can produce spurious features, can fail to distinguish between two nearby axons, and can fail to maintain continuity of axons through low-intensity signal regions. In each case, we manually annotated the corresponding tiles to mark false positives and false negatives as judged by a human observer. In the second and third rows of each respective figure, the *true-positives* are shown marked in cyan, *false-positives* are marked in yellow and *false-negatives* are marked in magenta. These examples visually illustrate that our proposed algorithm outperforms the baseline method in preserving the connectivity of the neurites and respecting their underlying tree structures. The baseline algorithm starts from a binarized likelihood and is consequently unable to skeletonize faint signals (see further zoomed-in regions (columns (b),(d)) in Figs. 3 and 4). In contrast, DM-skeleton utilizes the analog likelihood image for Morse-based skeletonization and is able to better preserve the continuity of the axon fragments.

We would like to note that our approach to analyzing tracer injection data is not conceptually tied to using the Discrete Morse algorithm for skeletonization of the shape, and other approaches to skeletonization could also be then used as input to the subsequent summarization step. We have found in our work that the DM based skeletonization approach produced good quality results for our application and outperformed a standard baseline, however if better skeletonization methods for this type of data become available in the future, the overall proposed workflow and approach to the analysis of such data types would still apply, while swapping out the DM module for skeletonization.

These visual observations are quantified in Table 1 and show that the proposed technique outperforms the baseline method. For the WSI data set, the F1 score is 0.97 for DM-skeleton (0.6 for bwskel) and the IOU score 0.94 for DM-skeleton (0.55 for bwskel) showing sufficiently high quality suitable for the desired scientific application of the

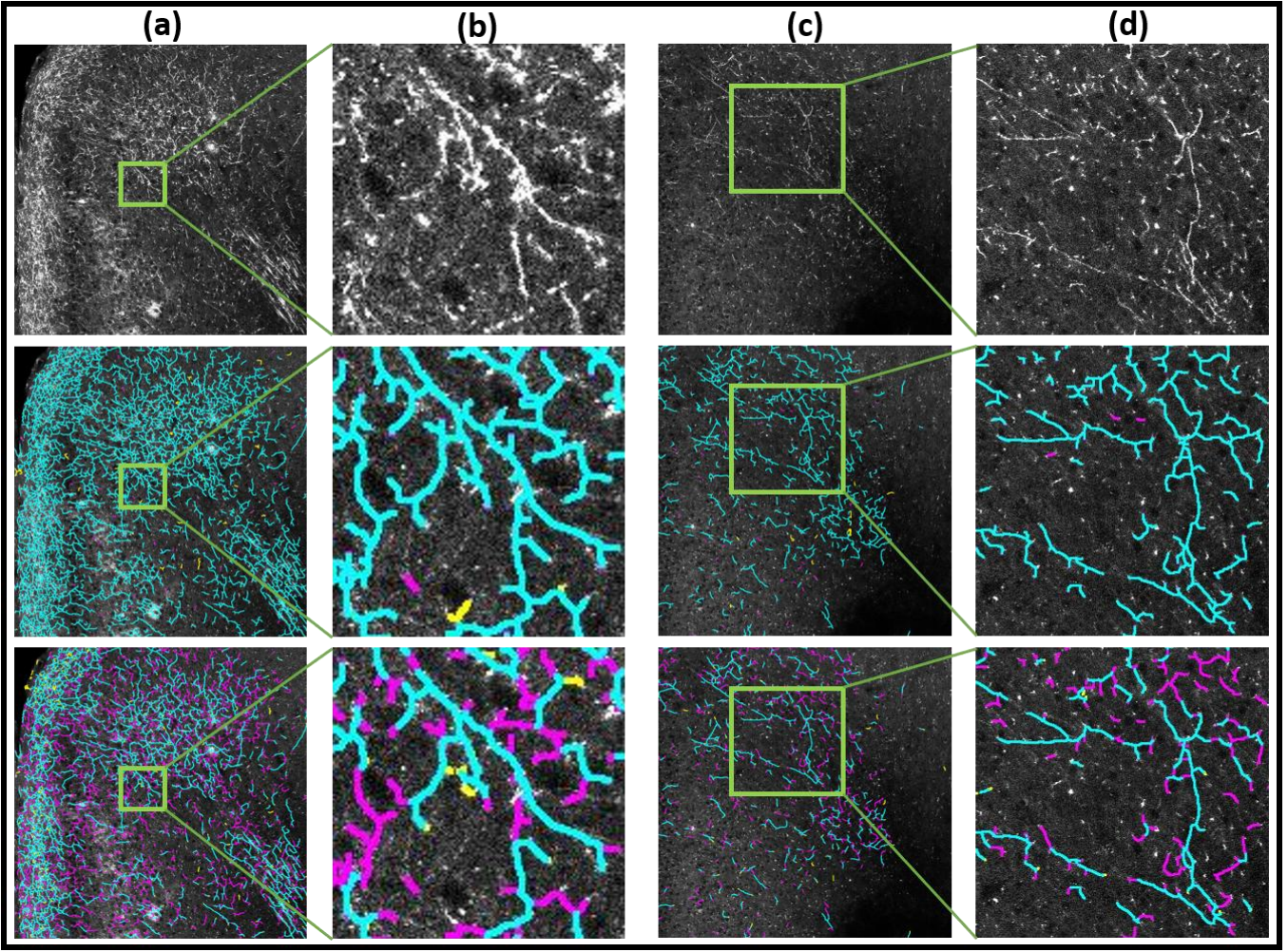

**Fig. 4 / Results of STP (serial two-photon images) neuron skeletonization.** Columns (a),(c) show image tiles and columns (b),(d) show zoom ins. Top row shows the original images, the middle row the results of DM-skeleton, and the bottom row the baseline results using bswkel. Axon fragments are seen in the original image as groups of lines of enhanced intensity. Results of visual-manual annotations for errors are also shown. *True-positives* are in *cyan*, *false-positives* are in *yellow* and *false-negatives* are in *magenta*. The results show better continuity and detection of the fragments by the DM-skeleton method over the baseline method.

|  | Precision | Recall | $F_1$-score | IOU |
|---|---|---|---|---|
| Proposed DM2D | 0.94 | 0.99 | 0.97 | 0.94 |
| MATLAB bwskel() | 0.87 | 0.60 | 0.71 | 0.55 |

(a) Table for comparison of techniques in the WSI dataset.

|  | Precision | Recall | $F_1$-score | IOU |
|---|---|---|---|---|
| Proposed DM2D | 0.92 | 0.96 | 0.94 | 0.89 |
| MATLAB bwskel() | 0.90 | 0.59 | 0.72 | 0.56 |

(b) Table for comparison of techniques in the STP dataset.

**Table 1**: The two tables shown below give the metrics for comparison of the proposed technique, **DM2D** with the baseline method, **MATLAB *bwskel()***. The metrics are calculated based on the *true-positive*, *false-positive*, and *false-negative* pixels in the detected image, corresponding to the manually annotated *Ground-Truth* image.

technique. The scores for the STP data set are only slightly lower (see 1.4 for further details). Note that in application of the methodology to whole brain data sets we encountered some spatially distinct compartments of overall poor performance due to tissue processing issues or imaging artifacts (folded sections and vasculature at the base of the brain for the WSI image data, and artifactually saturated islands of voxels at edges or the brain). These artifacts were visible in low resolution versions of the images, and the corresponding tissue regions were masked out. Also, the injection regions showed label saturation with individual axons not visible separately, as can be expected due to the dense labeling of processes and cells within the injection region. The injection regions were separately detected using a signal threshold and were excluded from the analysis.

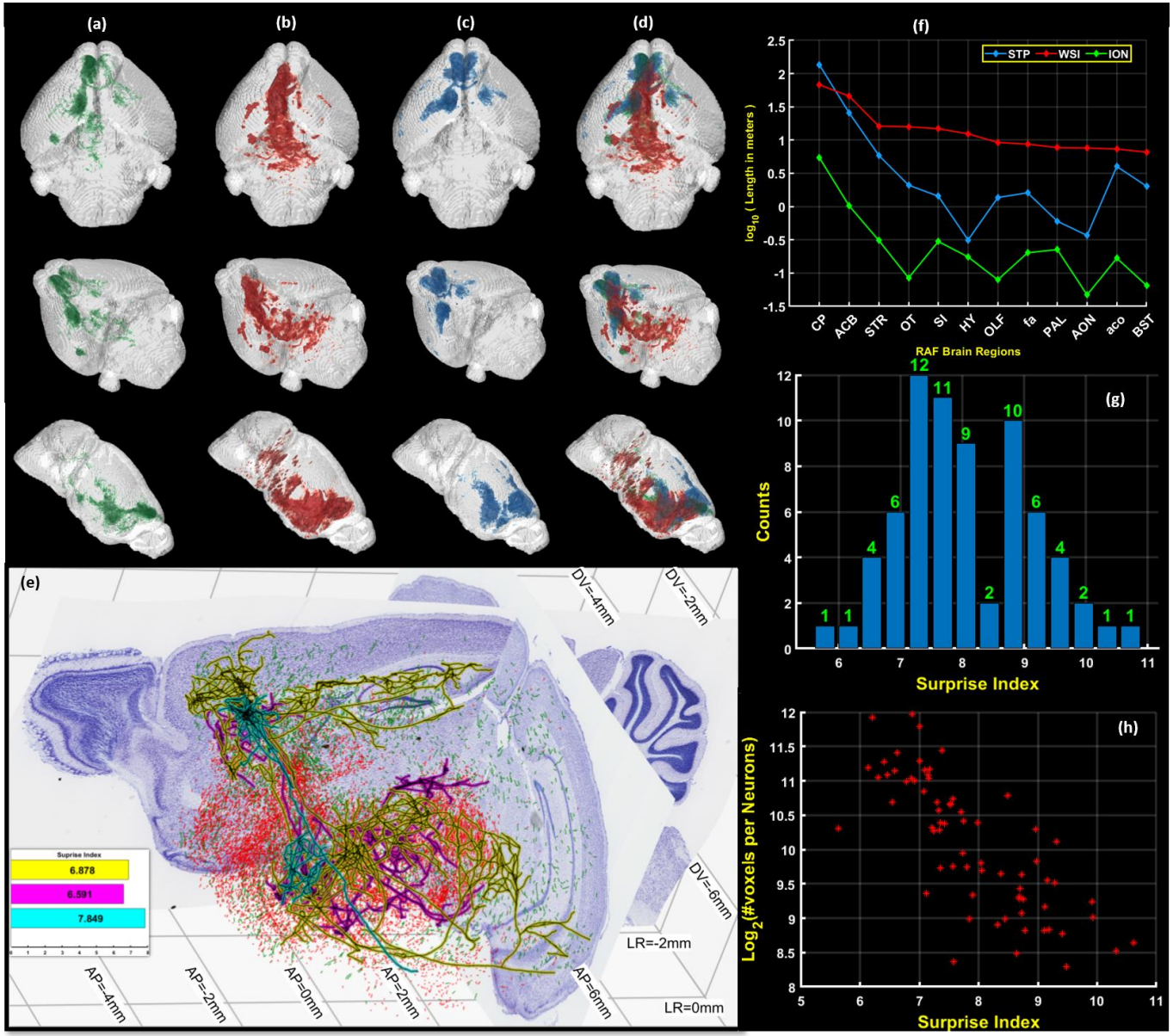

**Fig. 5 / 3D summarization of Skeletonized Data.** Columns **(a),(b),(c),(d)** show the line densities of the skeletons in the RAF space. **(a)** shows the density curated single neurons of the Mouselight-ION dataset. **(b)** shows the density of the line-fragments detected by DM2D for the WSI dataset. **(c)** shows the density of the line-fragments detected by DM2D for the STP dataset. **(d)** shows a comparison of the densities as depicted in *(a)-(c)*. **(e)** Whole neurons and fragments are shown from three different datasets, reconstructed in the space of a BAP-RAF atlas framework, all with injections in the Prelimbic area of the brain, to illustrate the relationship between single neuron reconstructions and tracer-labeled sets of axon fragments. The two Nissl-stained sections, one coronal and one sagittal are shown from the BAP-RAF mouse atlas together with atlas coordinate markings. Red fragments were reconstructed from a fluorescent WSI dataset and green fragments were reconstructed from an STP dataset, automatically annotated using the proposed algorithm. Only subset are shown to preserve visualization ability. Black lines with cyan, magenta, and yellow outlines show three example reconstructed neurons. A histogram of "surprise indices" of the set of 70 neurons is shown in the inset.**(f)** shows a graph showing the total lengths of the axon fragments from the tracer injected data sets as well as the lengths of the 70 single axons contained in 12 brain compartments, where the compartments were chosen by rank ordering the total length from the WSI injection in all brain compartments. The regions shown are *(CP:Caudoputamen; ACB:Nucleus accumbens; STR:Striatum; OT:Olfactory tubercle; SI:Substantia innominata; HY:Hypothalamus; OLF:Olfactory Areas; fa:corpus callosum, anterior forceps; PAL:Pallidum; AON:Anterior olfactory nucleus; aco:anterior commissure, olfactory limb; BST:Bed nuclei of the stria terminalis)* **(g)** shows a histogram of the surprise indices of the 70 individually reconstructed neurons when compared to the WSI dataset. **(h)** shows that the surprise index is negatively correlated with the log of the total length of the axons but this correlation is not very tight.

To visualize the results in the context of the whole mouse brain, we mapped the skeletonized axon fragments to the Brain Architecture Project mouse brain Reference Atlas Framework[31] for regional projection strength analysis.

**Summarization of detected axon fragments.** The output of the skeletonization step consists of tree-like graph fragments, corresponding to fragments of the underlying neurons projected onto the sectioning planes. This directly leads to the estimation of an areal density of the line fragments, with units of length per area (we use $\mu^2/mm$).

To convert into a volumetric density of length per unit volume one needs to divide by an estimated thickness of the physical or optical section from which these fragments are obtained. In case of the WSI image data, the brain was sectioned in the coronal plane with $20\mu$ section thickness, with alternating Nissl and Fluorescent Sections. We estimated the optical section thickness to be the full width at half maximumum of the Point Spread Function of the microscope (see Supplementary Sec. S.2) as $1.5\mu m$. Assuming an average axonal diameter of $0.5\mu m$, we assume that any fragment detected in the image plane represents an axon fragment of the same length within a slab of thickness of $2.5\mu m$. More sophisticated stereological corrections are possible for estimating the actual length of the axon fragment by assuming a distribution of orientations, but for the present purposes we keep to this crude estimate as a first approximation. Since no attempt has been made in the past to make the sort of length fragment estimation that we are performing, so that our work can be regarded as a first step. If another optical plane thickness estimate is used, it will multiply our volumetric line density estimates by a constant factor. Similarly, while we make no attempt to account for the variation in orientation of the fibers with respect to the sectioning plane, assuming an angular distribution of the fragments could give rise to an additional multiplicative factor. An overall multiplicative factor from either source will not affect the estimates of the relative distribution of lengths across compartments. We note that good correlation with the actual axon length has been obtained in simulations using simple projections of the axonal fragments onto sectioning planes[1].

An in plane estimate of the volumetric line density is then obtained by dividing the planar line density estimate (with units of length per unit area) by $2.5\mu m$. The WSI images are spaced $40\mu$ apart, so we further interpolate our line density estimate across the intervening $40/2.5 - 1 = 15$ optical sections to obtain a densely and uniformly sampled volumetric line density estimate across the brain, in the form of a volumetric line density associated with each spatial voxel. The interpolation effectively amounts to multiplication by a factor given by the ratio of the section spacing to the effective thickness of the optical plane.

The section images were mapped into a standardized reference atlas space (BAP-RAF) [31], and missing images interpolated, in order to obtain uniformly sampled volumetric densities in the BAP-RAF space (See Fig. 5(b) for 3D visualization of the volumetric line density data). Mappings were computed using our previously develoned Generative Diffeomorphic Mapping (GDM) framework for multimodal brain atlas mapping[32, 33]. Briefly, in this approach a target dataset (a series of 2D slices or a 3D volume) of a given contrast is generated using a sequence of transformations of the underlying reference brain, and parameters characterizing these transformations are optimized [34, 35] to minimize a discrepancy between the transformed reference brain and the target brain. These transformations are then applied to the coordinates of the points in each skeletonized fragment in order to map that fragment into the reference space for purposes of visualization as well as quantification. The transformations are also applied to the 2D volumetric line density images to map them into the 3D RAF space.

A similar approach was utilized for the STP data[36], where the sections are sampled optically at $50\mu m$ spacing without any missing images. The same method was applied to interpolate the line density data from the optical sectioning plane to intermediate non-sampled planes, and to subsequently to obtain volumetric line densities (See Fig. 5(c) for 3D visualization of Data). The FWHM was estimated for STP as as $2\mu m$ and the axon diameter as $0.5\mu m$, giving a total optical thickness of $3\mu m$.

In addition to the tracer injection data, we used a curated set of 70 single-neurons with somata in the Prelimbic area of cortex (PL) composed of 69 neurons drawn from from https://neuroxiv.org/, data originally collected in [37], and one neuron from the Mouselight data set[38]. These single neurons were then mapped to the BAP-RAF atlas and voxelized line densities for the collection of the 70 axons were computed (Fig. 5(c)) for comparison with the line densities derived from the tracer injection data.

A 3D visualization of the summarized density of the three datasets, is shown in Fig. 5(d), where green, red and blue represents the single axon, WSI and STP datasets respectively. The Fig. 5(e) shows 2D cutaway slices in reference planes showing the projections of the sampled neurons and line fragments mapped to the BAP-RAF atlas. The results shows a significant overlap of our detected fragments with the ION dataset. In the $20\mu m$ BAP-RAF atlas, we estimated the total line-lengths in different brain compartments by integrating the voxelized densities. We assumed ($50/3 - 1 \sim 15$) missing sections per imaged section for STP brain, since the inter optical-section spacing is $50\mu m$, and ($40/2.5 - 1 = 15$) missing sections per imaged section for WSI brain, where the inter-optical section spacing is $40\mu m$. We plot the logarithm of line-lengths in meters for 12 compartments in the left hemisphere of the brain in Fig. 5(f). The compartments were chosen so that the total line length

**A measure of "surprise" of single neurons over tracer injections.** Since tracer injections label a group of neurons with varying morphologies, individual axon reconstructions provide additional information not directly available from the tracer injection. Our density estimates allow a quantification of this additional information. We introduce an information theoretic "surprise" measure that quantifies the additional information provided by an axon reconstruction over a tracer injection using the relative entropy between the voxelized line density of a single axon and the voxelized line density of the tracer injection fragments.

The "surprise" measure is defined as follows: let the line densities of the axon fragments for a tracer injection, normalized by the total length, be given by $p_i$ in $i = 1..n$ voxels. Due to normalization, $p_i$'s are non-negative numbers which sum to 1, i.e. $\Sigma_i p_i = 1$, and can be interpreted as a probability density function. Note that a similar definition may be made at a compartment level, and also for retrogradely labeled somata from a retrograde tracer injection, which we will not explicitly write out here but which are simple generalizations. The numbers $p_i$ may be interpreted as probabilities of receiving a projection in the respective voxels or compartments for neurons with somata in the injection compartment.

The projections are composed of individual neurons, each of which project to the same set of voxels, but any individual neuron or neuronal type will not in general have non-zero density in each tracer-injection voxel. Consider a set of $m$ single neurons composing the tracer injection. For each of these neurons we empirically define a neuron-specific projection density $q_{ij}$ where $j = 1..m$ by dividing the length of the neuron contained in the $i^{th}$ voxel, by the total length of the neuron, with $\Sigma_i q_{ij} = 1$. Without loss of generality we assume that the voxels are equal in size. When considering unequally sized regions or voxels, one would compute a density in the region then normalize that density. i.e. $q$ is a left stochastic matrix.

The "surprise" $S_j$ for neuron $j$, as compared to the tracer injection $p_i$, is then defined to be the relative entropy of the two distributions (the Kullback-Leibler divergence):

$$S_j = \Sigma_i q_{ij} \log_2(q_{ij}/p_i) \tag{1}$$

.

Fig. 5(g) shows a histogram of $S_j$ for the set of 70 curated neurons with somata in PL as compared to the cell-type nonspecific tracer injection in PL. For this analysis we used a voxel size of $100\mu m$. It is to be noted, that the analysis was possible due to the mapping of all data sets to a common coordinate system provided by the reference atlas.

To gain an intuitive understanding of the surprise measure it is useful to consider some limiting cases. It is easy to prove that the most "surprising" morphological cell type given an average projection pattern $p_i$ would satisfy $q_{ij} = \delta_{ik}$ where $k = \arg\min_i p_i$. This corresponds to a single-neuron that projects only to the voxel $k$ with the weakest projection $p_k$, and has the corresponding surprise $S_{max} = -\ln(\min p_i)$. If all the $p_i$ were equal for $i = 1..n$ then $S_{max} = \log_2(n)$, however in general the projection densities will vary across voxels. Generally, we expect that the surprise would be larger for localized axons. This trend is indeed borne out in the scatter plot shown in Fig. 5(h), where a negative correlation is seen between the length of the axons and the surprise index, however this correlation is not very tight, since the tracer does not have uniform projection density in all voxels.

On the other hand, the least surprising cell type is one that has the same projection pattern as $p_i$, i.e. $q_{ij} = p_i$, and for this type the surprise is zero. Such a neuron could be considered "typical" in a probabilistic sense, given the tracer injection based projection density.

Fig. 5(e) shows three single neurons and the corresponding surprise indices. The estimation of the surprise measure has one technical subtlety - the measure assumes that the single neurons are constituents of the tracer injected neuron set, and thus only have nonzero density where the tracer injection density is nonzero. In practice however the single neurons cannot be traced in the same brain as the tracer injection, so due to statistical estimation issues, it may be the case that a voxel with nonzero density for a single axon has zero density from the tracer. To address this issue, we interpolated the estimated tracer injection density to the small number of voxels where the tracer injection did not produce any fragments, but the single axon did. To perform this interpolation we used the weighted Nearest Neighbour interpolation technique [39], which has been proven to be statistically consistent and therefore can be expected to have reasonable estimation performance.

It is to be noted that since we start with 2D image data, the axon fragments are projected onto a 2D plane. Since the axons are not always parallel to the imaging plane, this leads to an underestimate of the axon lengths. Therefore, the estimates obtained in our study should be regarded as a lower bound to the true biological axon lengths. This can be rectified by utilizing 3D volumetric imaging data at high resolution, with a 3D skeletonization step replacing the 2D skeletonization step. However, our current approach bring us closer to the biological reality of the underlying axons in comparison with the previous approach using only the fluorescent intensities of voxels. Moreover, in current studies involving tracer injections, the plane of section is consistent across experimental brain data sets, so that the axon fragment densities can be meaningfully interpreted as the projected axon length densities onto the corresponding sectioning planes (usually the coronal plane).

# 4 Discussion

In this manuscript we have introduced the usage of Discrete Morse techniques for the analysis of neuroanatomical data pertaining to brain circuit mapping, utilizing injections of tracer substances to label groups of axons projecting out of the injection site. This Topological Data Analysis approach has the advantage of being able to utilize non-local connectivity properties of the data, which led to robust performance in our application, by tracing labeled axon fragments through regions of low label intensity in noisy images.

We developed an approach (DM-skeleton) to the skeletonization of labeled axons in 2D microscopic image data using the Discrete Morse method combined with previously developed deep net methods utilizing TDA to provide likelihood maps from original image data. The relevant algorithms were codified into a computational pipeline which we provide in this manuscript together with data examples (see https://data.brainarchitectureproject.org/pages/skeletonization code packages). DM-skeleton showed good performance (F1 scores of 0.94 and 0.97 respectively on STP and WSI data) compared to a baseline skeletonization technique (*bwskel()* from MATLAB). We expect that the 2D skeletonization pipeline we provide will be applicable to other image data sets as well that contain line-like or tree-like objects.

We further introduced a new method for summarizing tracer injection data. The 2D fragments obtained from the DM-skeleton pipeline applied to tracer-injected brain image data were used to compute areal line densities (with units of length per unit area), which were then converted into a 3D volumetric density of length per unit volume by dividing by an appropriate optical section thickness as well as interpolation to account for missing optical sections. This provides a biologically meaningful quantification of the tracer injection data, in contrast with previous quantification using total fluorescent intensity or by counting fluorescently labeled voxels. Since the skeletonization step vectorizes high resolution microscopic image data, it also leads to very significant data compression while retaining biologically meaningful information and without loss of spatial resolution.

Topological Data Analysis methods are known to be computationally expensive compared with other methods. In our application the computational bottleneck comes from the persistence-guided discrete Morse-based framework. The computation of persistence pairings can have a worst-case time complexity of $O(n^3)$ (although it is usually significantly faster in practice), where $n$ is the number of cells that make up the input cell complex, which in our case corresponds to the number of pixels in the image tile input to the pipeline. To address this bottleneck, we utilized the DIPHA package [40] to compute persistence pairings. This is a distributed algorithm providing a significant speedup compared to centralized persistence algorithms. Further code optimizations over our current implementation are possible and run-time could be further reduced in future work. The current pipeline of *DM2D* which takes the likelihood produced by *DM++* overlayed with the binary mask takes ∼ 150 seconds for a STP brain section (∼ $11K \times 8K$ pixels), while it takes ∼ 200 seconds for a WSI section (∼ $22K \times 18K$ pixels). The post-processing step (in MTALAB) to convert the detected skeletons to vectorized *GeoJSONs* for web display takes ∼ 3 − 5 minutes per section. These estimates were made on an Intel Xeon Dual-CPU Quad-GPU (NVIDIA RTX 2080TI) machine with 512 GB of RAM. Despite the higher computational complexity, the conceptual elegance and theoretical transparency, performance improvement in detection of fragments, significant reduction in human proof-reading times and incorporation of prior biological structure are arguments in favor of the approach proposed here.

# References

[1] Rubio-Teves, M. *et al.* Benchmarking of tools for axon length measurement in individually-labeled projection neurons. *PLoS Computational Biology* **17**, e1009051 (2021).

[2] Banerjee, S. *et al.* Semantic segmentation of microscopic neuroanatomical data by combining topological priors with encoder–decoder deep networks. *Nature machine intelligence* **2**, 585–594 (2020).

[3] Dey, T. K. & Wang, Y. *Computational topology for data analysis* (Cambridge University Press, 2022).

[4] Edelsbrunner, H. & Harer, J. *Computational Topology : an Introduction* (American Mathematical Society, 2010).

[5] Carlsson, G. Topology and data. *Bull. Amer. Math. Soc.* **46**, 255–308 (2009).

[6] Chazal, F. & Michel, B. An introduction to Topological Data Analysis: fundamental and practical aspects for data scientists. *CORR* (2017).

[7] Lum, P. Y. *et al.* Extracting insights from the shape of complex data using topology. *Scientific Reports* **3** (2013).

[8] Tierny, J. *Topological Data Analysis for Scientific Visualization* (Springer, 2018).

[9] Edelsbrunner, Letscher & Zomorodian. Topological persistence and simplification. *Discrete & Computational Geometry* **28**, 511–533 (2002).

[10] Buchet, M., Hiraoka, Y. & Obayashi, I. *Persistent Homology and Materials Informatics*, 75–95 (Springer Singapore, Singapore, 2018).

[11] Singh, G. *et al.* Topological analysis of population activity in visual cortex. *Journal of vision* **8**, 11 (2008).

[12] Platt, D. E., Basu, S., Zalloua, P. A. & Parida, L. Characterizing redescriptions using persistent homology to isolate genetic pathways contributing to pathogenesis. *BMC Systems Biology* **10**, S10 (2016).

[13] Lamar-León, J., García-Reyes, E. B. & Gonzalez-Diaz, R. Alvarez, L., Mejail, M., Gomez, L. & Jacobo, J. (eds) *Human gait identification using persistent homology.* (eds Alvarez, L., Mejail, M., Gomez, L. & Jacobo, J.) *Progress in Pattern Recognition, Image Analysis, Computer Vision, and Applications*, 244–251 (Springer Berlin Heidelberg, Berlin, Heidelberg, 2012).

[14] Lee, Y. *et al.* Quantifying similarity of pore-geometry in nanoporous materials. *Nature Communication* 15396 (2017).

[15] Li, Y., Wang, D., Ascoli, G. A., Mitra, P. & Wang, Y. Metrics for comparing neuronal tree shapes based on persistent homology. *PloS one* **12** (2017).

[16] Chaudhuri, R., Gerçek, B., Pandey, B., Peyrache, A. & Fiete, I. The intrinsic attractor manifold and population dynamics of a canonical cognitive circuit across waking and sleep. *Nature Neuroscience* **22**, 1512–1520 (2019).

[17] Kanari, L. *et al.* A topological representation of branching neuronal morphologies. *Neuroinformatics* **16**, 3–13 (2018).

[18] Delgado-Friedrichs, O., Robins, V. & Sheppard, A. Skeletonization and partitioning of digital images using discrete morse theory. *IEEE Trans. Pattern Anal. Machine Intelligence* **37**, 654–666 (2015).

[19] Gyulassy, A. *et al.* Topologically clean distance fields. *IEEE Trans. Visualization Computer Graphics* **13**, 1432–1439 (2007).

[20] Robins, V., Wood, P. J. & Sheppard, A. P. Theory and algorithms for constructing discrete morse complexes from grayscale digital images. *IEEE Trans. Pattern Anal. Machine Intelligence* **33**, 1646–1658 (2011).

[21] Sousbie, T. The persistent cosmic web and its filamentary structure – I. Theory and implementation. *Monthly Notices of the Royal Astronomical Society* **414**, 350–383 (2011).

[22] Dey, T. K., Wang, J. & Wang, Y. *Graph reconstruction by discrete morse theory*, 31:1–31:15 (2018).

[23] Forman, R. Morse theory for cell complexes. *Advances in Mathematics* **134**, 90 – 145 (1998).

[24] Wang, S., Wang, Y. & Li, Y. *Efficient map reconstruction and augmentation via topological methods*, SIGSPATIAL '15, 25:1–25:10 (ACM, New York, NY, USA, 2015).

[25] Dey, T. K., Wang, J. & Wang, Y. *Road network reconstruction from satellite images with machine learning supported by topological methods* (2019). To appear.

[26] Fornito, A., Zalesky, A. & Bullmore, E. T. in *Chapter 3 - connectivity matrices and brain graphs* 89 – 113 (Academic Press, San Diego, 2016).

[27] Lin, M. K. *et al.* A high-throughput neurohistological pipeline for brain-wide mesoscale connectivity mapping of the common marmoset. *Elife* **8**, e40042 (2019).

[28] Ragan, T. *et al.* Serial two-photon tomography for automated ex vivo mouse brain imaging. *Nature methods* **9**, 255–258 (2012).

[29] Matho, K. S. *et al.* Genetic dissection of the glutamatergic neuron system in cerebral cortex. *Nature* **598**, 182–187 (2021).

[30] Josh Huang, Z. & Zeng, H. Genetic approaches to neural circuits in the mouse. *Annual Review of Neuroscience* **36**, 183–215 (2013).

[31] Tward, D. J. *et al.* 3d multimodal histological atlas and coordinate framework for the mouse brain and head. *Nature* (2025, Under review).

[32] Tward, D. *et al.* Diffeomorphic registration with intensity transformation and missing data: Application to 3d digital pathology of alzheimer's disease. *Frontiers in neuroscience* **14**, 52 (2020).

[33] Tward, D. J. *et al.* Solving the where problem and quantifying geometric variation in neuroanatomy using generative diffeomorphic mapping. *bioRxiv* (2024).

[34] Beg, M. F., Miller, M. I., Trouvé, A. & Younes, L. Computing large deformation metric mappings via geodesic flows of diffeomorphisms. *International journal of computer vision* **61**, 139–157 (2005).

[35] Tward, D. J. An optical flow based left-invariant metric for natural gradient descent in affine image registration. *Frontiers in Applied Mathematics and Statistics* **7**, 718607 (2021).

[36] Kim, Y. *et al.* Brain-wide maps reveal stereotyped cell-type-based cortical architecture and subcortical sexual dimorphism. *Cell* **171**, 456–469 (2017).

[37] Gao, L. *et al.* Single-neuron projectome of mouse prefrontal cortex. *Nature neuroscience* **25**, 515–529 (2022).

[38] Winnubst, J. *et al.* Reconstruction of 1,000 projection neurons reveals new cell types and organization of long-range connectivity in the mouse brain. *Cell* **179**, 268–281 (2019).

[39] Belkin, M., Hsu, D. J. & Mitra, P. Overfitting or perfect fitting? risk bounds for classification and regression rules that interpolate. *Advances in neural information processing systems* **31** (2018).

[40] Bauer, U., Kerber, M. & Reininghaus, J. *Distributed Computation of Persistent Homology*, 31–38 (2014).

[41] Ragan, T. *et al.* Serial two-photon tomography for automated ex vivo mouse brain imaging. *Nature Methods* **9**, 255–258 (2012).

[42] Pinskiy, V. *et al.* A low-cost technique to cryo-protect and freeze rodent brains, precisely aligned to stereotaxic coordinates for whole-brain cryosectioning. *Journal of neuroscience methods* **218**, 206–213 (2013).

[43] Pinskiy, V. *et al.* High-throughput method of whole-brain sectioning, using the tape-transfer technique. *PloS one* **10** (2015).

[44] Dey, T. K., Wang, J. & Wang, Y. *Improved road network reconstruction using discrete morse theory*, 58–66 (2017).

[45] Dempster, A. P., Laird, N. M. & Rubin, D. B. Maximum likelihood from incomplete data via the em algorithm. *Journal of the royal statistical society: series B (methodological)* **39**, 1–22 (1977).

[46] Milnor, J. W. *Morse Theory* 5th edn. Annals of Mathematics Studies (Princeton University Press, 1973).

[47] Forman, R. A user's guide to discrete Morse theory. *Séminare Lotharinen de Combinatore* **48** (2002).

[48] Zomorodian, A. J. *Topology for Computing* Cambridge Monographs on Applied and Computational Mathematics (Cambridge University Press, 2005).

[49] Edelsbrunner, H. & Harer, J. Persistent homology – a survey.

[50] Chung, M. K., Bubenik, P. & Kim, P. T. Prince, J. L., Pham, D. L. & Myers, K. J. (eds) *Persistence diagrams of cortical surface data.* (eds Prince, J. L., Pham, D. L. & Myers, K. J.) *Information Processing in Medical Imaging*, 386–397 (Springer Berlin Heidelberg, Berlin, Heidelberg, 2009).

## Data and code availability

## Acknowledgements

This work is in part supported by National Science Foundation under grants CCF-1740761, RI-1815697, DMS-1547357 and CCF-2112665, National Institute of Health under grant R01-EB022899, MH114821, MH114824, NS121761 and

NS132173. We would also like to thank the Crick-Clay Professorship and the Mathers Charitable Foundation for support to the Brain Architecture Project.

The authors thank Lucas Magee for his help in initial stages of the analysis and for advising CS in later stages. The authors thank Max Richman, Somesh Balani, and Patrick Flannery, for their help in annotating the brains to generate the ground-truth data for analysis and for proofreading the detects. We would also like to thank Linus Manubens-Gil for his help in manually curating the single neuron dataset. We thank Pratik Purohit and Ken Arima for their help in streamlining the computational pipeline.

## Author contributions

PM proposed applying TDA methods to neuron skeletonization and conceived the study together with YW. SB, CS, YW and PM iteratively designed the computational pipeline for DM, which was implemented and tested by SB and CS. DJT carried out atlas registration data visualizations in atlas space. SS acquired the WSI dataset. PM developed the single neuron surprise measure. SB and DJT applied the computational pipelines to the brain data, and together with YW and PM analyzed the results. SB, YW and PM wrote the manuscript with input from DJT and SS.

## Competing interests

The authors declare no competing interests.

# Methods

## 1.1 Data Collection.

Tract tracing is the gold standard for studying mesoscale axonal projections in vertebrate brains. Each anterograde tracer injection can label hundreds to thousands of neurons. The fluorescent label from the tracer fills the axons, showing the neuronal projection pattern across the whole brain.

The Serial Two-Photon (STP) dataset presented in this paper was collected as a part of Brain Initiative Cell Census Network [29]. Cre-dependent transgenic mouse lines were crossed with IslFlp reporter lines. Flp-dependent AAV tracers were utilized to reveal cell type-specific axon connection [30]. Each brain was prepared and imaged using STP tomography [41] with 1µm × 1µm in-plane resolution, and sectioned coronally every 50 µm. Two channels of 16-bit data were collected, where Channel 1 collected the autofluorescence and Channel 2 collected the fluorescent tracer information. Only Channel 2 data were used in the subsequent analysis. One STP dataset was involved in the development and demonstration of methods in this paper (available from: ftp://download.brainimagelibrary.org:8811/biccn/huang/connectivity/anterograde/190322_JH_HK0126_PlexinD1LSLflp_PL_male_processed/). The Data presented in this paper includes a dataset with injection in prelimbic region of the brain. The visualization of the STP dataset used in the paper can be viewed from https://data.brainarchitectureproject.org/pages/skeletonization.

The fluorescent Whole-slide Imaging (WSI) dataset presented in this paper was collected as a part of Mouse Brain Architecture (MBA) project, where fluorescent tracers were injected into the same brain. C57BL/6J mice were acquired from Jackson Laboratories (stock 000664) under IRB protocol #498813-28 according to protocols approved by the Animal Care and Use Committee at Cold Spring Harbor Laboratory. Two tracer injections were placed in the right hemisphere of each mouse, both anterograde (AAV2/1.CAG.tdTomato.WPRE/SV40, AAV2.1CB7.CI.EGFP.WPRE.RBG). Approximately 2.3nl of virus was injected using a Nanoject II injection system before a 4 week incubation period. All samples were histologically processed using methods previously described [27, 42, 43]. The brain was fixed, embedded in freezing agent, and serially cut at 20µm using the tape-transfer method to minimize tissue distortion [42, 43]. All slides were scanned by a Nanozoomer 2.0HT with a 20x objective (0.46µ m in-plane resolution) and saved in an uncompressed RAW format. Image cropping, conversion and compression to per section *JPEG-2000* files were performed. Alternating sections were imaged with either widefield imaging after Nissl staining or fluorescent imaging at 0.46µm × 0.46µm in-plane resolution. All images were recorded with 3 (RGB) channels with 12-bit data in each channel. For the purposes of this manuscript, we only use the AAV2/1.CAG.tdTomato.WPRE/SV40 injection in the prelimbic area of the brain. The microscopic images of the WSI dataset used in the paper can be viewed from https://data.brainarchitectureproject.org/pages/skeletonization.

## 1.2 Data pre-processing.

The WSI dataset was registered in 3D using Nissl sections. Fluorescent sections were subsequently cross-registered to the adjacent Nissl sections and formed a 3D volume [33]. Both the STP and WSI datasets were first processed with fluorescent labeled axon signal detection [2]. The original images composed 1µm × 1µm pixels for STP Coronal sections spacing was 50µm for STP and 40µm for WSI.

The STP dataset was first processed with a combined TDA and deep net based method [2] for the detection of the tracers. The network, termed as DM++, takes in whole STP sections and divides them into 512 × 512 pixel tiles. These tiles are passed through a TDA stage based on Discrete Morse [44] and a CNN stage for determining the topological and axonal priors, respectively. The topological priors capture the faint connectivity which is used to boost the performance of the CNN in a supervised Siamese network using the dual priors that comprise the DM++ framework. The final output likelihood map is converted into a binary mask for the neuronal processes using an optimal empirically determined threshold. This captures most of the processes in the tiles, which are then stitched back together to form a mask for an entire reference section of the brain.

The preliminary outputs of process detection were manually verified for the entire brain by a histotechnologist using MATLAB. Briefly, the preliminary outputs consisting of detected signal were masked with the original brain section image and error corrected using a MATLAB-based pixel annotation tool (refer Section 1.5 for further details). The filled processes were identified as those having a brighter intensity compared to the background. The proofread brain from the previous step was annotated in the format of binary images. The images were further downsampled to the desired resolution by summing pixels appropriately.

## 1.3 DM-Skeleton

This section provides details for the DM-Skeleton pipeline corresponding to the workflow in Fig. 1.

### 1.3.1 Step 1: Pre-processing

Each image volume is loaded as an image stack, and converted into a density field $\rho : K \to \mathbb{R}$ defined on the 2D-cubical complex grid $K$, where each vertex corresponds to a pixel in the input image and has a density value. For whole-brain tracer injection STP and PMD data, a significant portion of the raw images is background (see the example in Fig.

1). Hence we first applied the learning-based process-detection module [2] to remove the background and segment the foreground consisting of labeled processes. See the previous section for more details. The resulting foreground is segmented and binary-masked. We further apply a Gaussian filter to smooth the values across the domain.

### 1.3.2 Step 2: Skeletonization.

The Discrete Morse graph reconstruction algorithm [22, 24] takes a density field as input and outputs a graph skeleton capturing center-lines passing through relatively high density regions. In our case, the input density field $\rho : \mathsf{K} \to \mathbb{R}$ is defined at vertices of a 2D-cubical complex $\mathsf{K}$ of the domain, which is a collection of squares (2-cells), their edges (1-cells), and vertices (0-cells), forming a planar structure. In all subsequent operations, only the 2-skeleton of this 2D-cubical complex $\mathsf{K}$ is needed, that is, we assume $\mathsf{K}$ consists of vertices, edges, and squares.

To explain the main idea, consider first the smooth case where we have a smooth function $\rho : \Omega \to \mathbb{R}$ over the domain $\Omega$. Consider the terrain of the density function values plotted over the domain (Fig. 2) where the terrain of a function defined on $\mathbb{R}^2$ is given. The underlying graph skeleton of $\rho$ can be captured by the paths connecting the peaks on the mountain ridges through the intervening saddle points (Fig. 2a). These paths form the so-called *1-unstable manifold* in Morse theory, and are defined by the integral lines "connecting" saddle points to local maxima (Fig. 2b). An integral line is a curve in the domain where at any point on it, its tangent vector coincides with the gradient of the density field. Integral lines are thus intuitively flow lines, following the steepest descending direction of the density fields.

Inside the algorithm, roughly speaking, ridges (as defined by pairs of (saddle, maximum)) are associated with certain persistence values, as quantified by the so-called persistent homology [9]. The persistence values can be interpreted as importance scores. This makes it possible to filter out ridges of "low importance", which are assumed to be associated with noise, from the final output by providing the algorithm with a persistence threshold. An example of simplification for a very simple 1D function is shown in Fig. 2c and 2d.

For the input to our algorithm, we have a density field $\rho : \mathsf{K} \to \mathbb{R}$ defined at vertices of a 2D-cubical complex $\mathsf{K}$ of the domain $\Omega$ (a 2D region). Following [22], discrete Morse theory [23] is used to capture the mountain ridges mentioned above, combined with the persistence algorithm to measure importance. See Supplementary Materials for a short tutorial on the DM-algorithm. To improve the efficiency of the algorithm, we modified the algorithm of [22] so that it works directly with 2D-cubical complexes and also uses DIPHA [40] to compute persistence pairs in a distributed manner.

These mountain ridges cover the axonal branches as locally, points in the image along these branches tend to have relatively higher density (signal strength) than off the branches. The global nature of the 1-stable manifolds makes the output skeleton robust to small gaps in signal, and effective at capturing junctions; see e.g., Fig. 2a, where the global nature of 1-manifolds connects through low-density region around the Y-junction.

In ideal circumstances, we would find a persistence threshold that would remove all of the noise and only keep the ridges that make up the true neuron tree. However, because of the noisy nature of biological data and also the Discrete Morse graph reconstruction algorithm will not necessarily output a tree, we cannot take the algorithm's output as a final output. Instead, we first run the algorithm with a low persistence threshold such that we do not remove any ridges that would be part of an ideal output. Then we simplify the Morse graph skeleton in the next step.

### 1.3.3 Step 3: Simplification using a "Haircut" step.

The output of the above persistence-guided Morse-based framework is a geometric graph $G$, also referred to as the Morse graph skeleton. The morse graph $G$ has a set of trees in them for each connected component in the likelihood image. These trees provide a good initial estimate of the skeletons. However, the likelihood images produced from the convolution-based framework of Process Detection [2], have a few false detects. We use the binary threshold on likelihood image, from the *DM++* algorithm to mask out most of the false detects to obtain a modified graph $G'$. The modified morse graph $G'$ also can have several small side branches, originating in paths connecting the real signal to nearby noise maxima. These small branches (~ 10 pixels) are characterized by path from the branch points on the graph (degree-2 nodes) to the endpoints on the graph (degree-1 nodes) that change direction at most once. $G'$ is pruned of these paths using this *Haircut* mechanism to produce a simplified Morse Graph $G_s$.

### 1.3.4 Summarization of the length of the Tracer Fragments

Spatial mappings (i.e. a 3D displacement vector at each voxel) were computed between our atlas and each target dataset using our generative diffeomorphic mapping framework previously developed and validated in human [32] and in mouse [33]. In this framework a target dataset is generated from a sequence of transformations of the atlas dataset. This sequence includes a diffeomorphism computed within the Large Deformation Diffeomorphic Metric Mapping framework [34] encoding changes in shape; a 12 parameter affine transformation encoding changes in scale, orientation, and position; a sequence of 3 parameter 2D rigid transforms (one per slice, only when a 2D serial section dataset is used); and a polynomial change of contrast to account for multimodality data. To account for missing tissue or artifacts, our procedure includes an Expectation Maximization algorithm [45] to compute the posterior probability that a given voxel in our target is good quality. Parameters characterizing these transforms are jointly optimized in a maximum a posteriori framework. Those characterizing the diffeomorphism are updated using Hilbert gradient descent [34], those characterizing the affine and rigid transforms are updated using Riemannian gradient descent [35], and those characterizing the contrast differences are updated by solving a weighted least squares problem. Once computed, each fragment was transformed into the coordinates of our atlas by applying the inverse transformation to each vertex, and leaving the edges (connectivity information) unchanged. A density with units of fragment length per unit volume was computed by assigning each line segment in 2D to the pixel where its center lay, and incrementing the density value at this pixel by the length of the line segment divided by the area of the 2D voxel (starting from 0), giving units of length per unit area. This was further divided by the optical thickness of the section to give units of length per unit volume as described in the main text.

### 1.3.5 Additional features.

The simplified Morse Graph $G_s$ was divided into skeletal-fragments for vectorization. For each connected component (>) in the detected image, a graph was constructed. The graph consisted of branchpoints, $B_p$ (nodes with degree > 2) and endpoints, $E_p$ (nodes with degree 1), collectively called as Critical Points $C_p$. Each skeletal fragment consists of path between any pair of connected $C_p$'s. Using a modified Depth-First Search, we walk through all the edges of the graph, to produce these Skeletal fragments. These fragments are vectorized as line-strings, with their length also recorded. The vectorized line strings are converted to geojsons, for display on the web (visit https://data.brainarchitectureproject.org/pages/skeletonization for more details).

## 1.4 Evaluation metrics: Precision, Recall and F1-score for evaluating results.

For skeletonized data, a set of manually annotated sample tiles were provided as ground truth. The precision and recall metrics calculated for evaluating skeletonization results depend on True Positives (TP), False Positives (FP) and False Negatives (FN). TP is calculated as the each pixel belonging to the detected image which is within a ceratin radius $D_r$ of the ground truth (GT) pixel ($D_r$ = 5 for WSI data; $D_r$ = 3 for STP data, empirically determined), where a GT pixel can contribute only once to the calculation to the metric. This prevents two neighboring lines from contributing to a single GT pixel. Pixels along a line do not necessarily exclusively contribute to the TP. All the pixels in the evaluated image which are not in the TP set and not in the GT set, are considered as FP, while all pixels in the GT set which are not in the TP set and not in the detected pixel set, are considered as FN. The second and third rows of the Figs. 3 and 4 shows the TP in cyan, FP in yellow and FN in magenta. For the second row, the evaluated image is the skeletonized image by our proposed method, **DM2D**, and the third row shows the detected image by the baseline technique, *bwskel()* from MATLAB.

Precision and recall are then routinely computed as:

$$Precision = \frac{TP}{TP + FP} \quad (2)$$

$$Recall = \frac{TP}{TP + FN} \quad (3)$$

The F1-score is the harmonic mean of precision and recall, i.e.,

$$F_1 = \frac{2 \cdot Precision \cdot Recall}{Precision + Recall} \quad (4)$$

The parameter *IOU*-score was calculated as binary classification metric,

$$IOU = \frac{TP}{FP + FP + FN} \quad (5)$$

## 1.5 Manual Annotation Tool.

The *DM2D* algorithm generally produced good detects, in almost all the regions except the injection regions in the brain, where performance is disrupted by image saturation effects. There are also a few easily identifiable regions of false detects around the edges of the brain tissue, and the blood vessels, where it the autofluorescence signal not originating from neuronal processes leads to false detects. These regions were manually corrected. The injection region was manually demarcated and eliminated form the skeletons. We used an in-house Matlab-based Annotation tool. To generate the GT image, we manually added the line-strings corresponding to the missed detects on the image. For removal of the systemic errors and the Injection region, we use a polygon-based deletion tool, as shown in the Extended Fig. 1, where every pixel within the polygon is deleted.

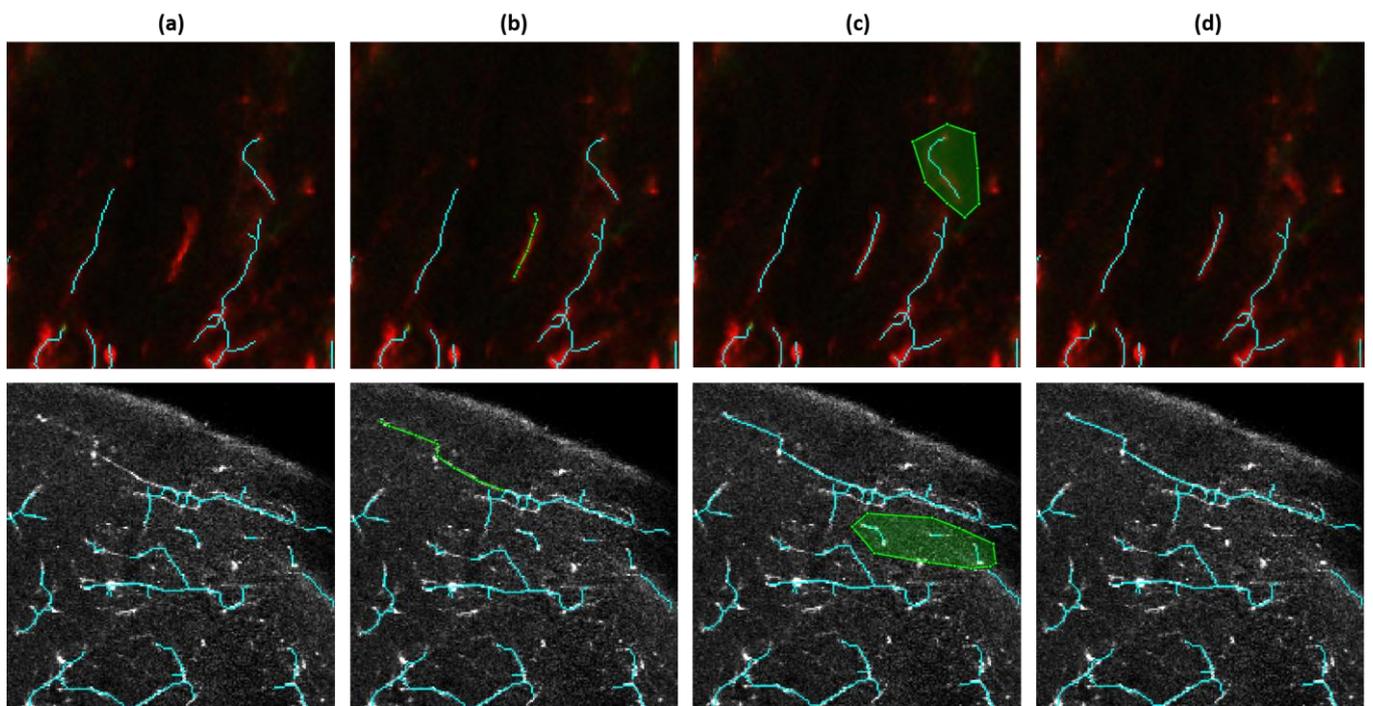

**Extended Fig. 1 / Manual Annotation Tool.** The MATLAB-tool used for generation of ground-truth, injection region removal, and correction of systematic errors at blood vessels and the brain outline. Column *(a)* shows a zoomed in portion of the contrast enhanced original image with the detected annotations overlayed in cyan, column *(b)* shows the annotator marking the missed neuron fragment in green, column *(c)* shows the annotator marking the *false-positives* using an arbitrary polygon, and the column *(d)* shows the result of the annotation after the correction corresponding to the images in column *(a)*. Each column shows an example of WSI image annotation in the top row, while the STP annotation is shown in the bottom image.

# Supplementary Materials

## S.1 DM Tutorial

### S.1.1 Morse theory and 1-stable manifolds in the smooth case

We first provide an informal description of the relevant part of Morse theory that motivates the graph reconstruction algorithm we use. Let $f : \mathbb{R}^d \to \mathbb{R}$ be a smooth function on $d$-dimensional Euclidean space. In our applications for processing neuron images, the domain is either $\mathbb{R}^2$ for 2D images or $\mathbb{R}^3$ for 3D volumetric image data. The gradient vector at a point $p \in \mathbb{R}^d$ is defined as [1]:

$$\nabla f(p) = -[\frac{\partial f}{\partial x_1}, \frac{\partial f}{\partial x_2}, \ldots, \frac{\partial f}{\partial x_d}]^T,$$

where $(x_1, \ldots, x_d)$ represents an orthonormal coordinate system for $\mathbb{R}^d$. In simple terms, $\nabla f(p)$ represents the *steepest descending direction* along which the function $f$ decreases fastest when moving away from $x$, and its norm $\|\nabla f(p)\|$ is the rate of this change. Gradient vectors for all points in $\mathbb{R}^d$ form a vector field on $\mathbb{R}^d$, called the *gradient vector field*. A point p with vanishing gradient; that is, $\nabla f(p) = [0, 0, \ldots, 0]^T$, is called a *critical point*; otherwise, it is a regular point. A critical point p of $f$ is *non-degenerate* if the Hessian matrix (formed by all second-order partial derivatives $[\frac{\partial^2 f}{\partial x_i \partial x_j}]$) has full rank; otherwise, it is degenerate.

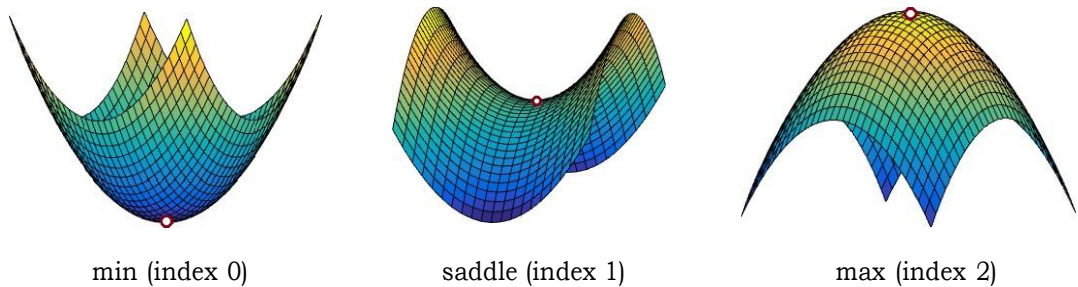

min (index 0)   saddle (index 1)   max (index 2)

**Supplementary Fig. 1** / Three types of critical points (of index 0, 1, and 2) for a Morse function defined on $\mathbb{R}^2$.

A *Morse function* is a smooth function where all critical points are (i) non-degenerate and (ii) have distinct function values. Morse functions are well-behaved functions whose critical points also have nice characterizations. For example, by Morse Lemma [46], for a Morse function $f$ on $\mathbb{R}^d$, there are $d + 1$ types of critical points, local minima (index 0), local maxima (index $d$), and $(d - 1)$-types saddle points (of indices from 1 to $d - 1$). In the case of a 2D Morse function $f : \mathbb{R}^2 \to \mathbb{R}$, there are three types critical points, local minima, local maxima, and saddle points as shown in Fig. 1. Critical points capture local behavior of the function $f$. The global variation can be partially captured by concepts such as integral lines and (un)stable manifolds. In particular, an *integral line* of $f$ is a maximal path in $\mathbb{R}^d$ such that the tangent vector of this path at any point coincides with the gradient $\nabla f(x)$. Intuitively, imagine that we view the graph of the function $f : \mathbb{R}^d \to \mathbb{R}$ as a "terrain" defined in $\mathbb{R}^{d+1}$; see Fig. 2 for an illustration where the last coordinate (i.e, the height of each point) corresponds to $f(x)$ at each $x \in \mathbb{R}^d$. The lift of an integral line to the surface of terrain is a "flow line" on the terrain that a water drop will follow when flowing in the direction of the steepest descending direction at any point. See the green solid curve in Fig. 2 for an example of a flow line, which starts at maximum $t_1$ and ends at minimum $v$.

Consider a flow line. The water drop will keep flowing till it reaches a point where there is no descending direction – these are exactly points where gradients vanishes, namely critical points. Hence flow lines (thus integral lines) have to "start" and "end" at critical points. The *unstable manifold* of a critical point is the union of points along all integral lines that ultimately "end" at this point. We are particularly interested in the 1-unstable manifolds, which are those flow lines that end at index $d - 1$ saddle points. In the example of a function defined in $\mathbb{R}^2$, these are pieces of curves that connecting maxima with saddles; see the white dotted curves in Fig. 2.

Such 1-unstable manifolds for saddles may be conceptualized of as "mountain ridges" of this terrain (graph of function $f$), connecting mountain peaks (maxima) to peaks (maxima) via saddles, and separating different valleys (basins of minima). There is also a dual concept of 1-*stable manifolds*, that correspond to "valley ridges", connecting minima to minima via saddles and separating mountain peaks.

---

[1]Note that this is a negative version of the sign convention used in classical Morse theory. We use this negated version as it is then aligned with the steepest descending direction of $f$; while the usual notion of the gradient vector indicates the steepest ascending direction.

In our setting, we can view a 2D image or a 3D image as a real valued density field on $\mathbb{R}^2$ or $\mathbb{R}^d$, respectively. Consider the terrain (graph) of this density field. The mountain ridges of this terrain can capture where strong signals of cell processes are. We will aim to use the 1-unstable manifold of this density function to capture its mountain ridges and further to capture the neuronal cell processes.

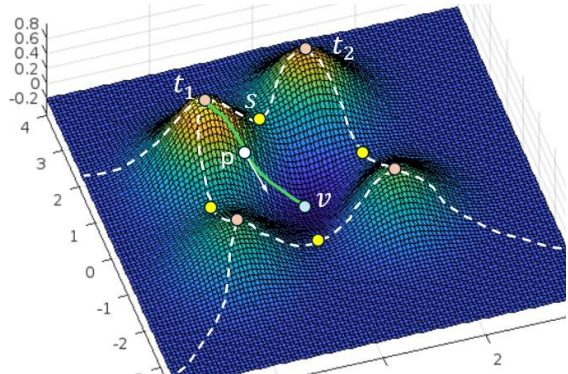

**Supplementary Fig. 2 /** White dotted lines are the union of 1-unstable manifolds. The green curve is an example of a flow line. Pink points are maxima, yellow points are minima, while the blue point $v$ is a minimum.

## S.1.2 Discrete Morse theory

Numerical data on a computer are discretely sampled and do not equal the mathematical notion of a smooth function. We can view the sampled 2D or 3D image data that as a discretization of a smooth function $\rho$ defined on $\mathbb{R}^2$ or $\mathbb{R}^3$. While one could compute the 1-unstable manifolds from a continuous extension of the discrete evaluations of $\rho$ at pixels (e.g, a piecewise-linear approximation), this could to be sensitive to approximation and numerical error as (un)stable manifolds are defined based on gradients. Simplifying / denoising the resulting (un)stable manifolds could be challenging.

Discrete Morse theory, a combinatorial analog of the classical Morse theory for cell complexes, was proposed by Forman[23, 47] as a mathematically well defined but explicitly discrete and computationally viable approach to Morse theory suitable for analysis of real life data sets.

We provide a brief informal description of some concepts from discrete Morse theory that are relevant to the present DM-based graph reconstruction algorithm. We utilize simplicial complexes, which are complexes consisting of building blocks called simplices, which in 2D consist of points, line segments, triangles and pyramids, glued appropriately along their faces. A $d$-dimensional (geometric) simplex is the convex combination of $d+1$ affinely independent points. Thus 0, 1, and 2-simplices correspond to vertices, edges, and triangles. See Fig. 3 for an example of a 2-dimensional simplicial complex, triangulating a square, consisting of a collection of vertices, edges, and triangles. In our applications, we use cubical complexes instead of simplicial complexes to represent images, where pixels are vertices, and cells are squares instead of triangles. However, we will simplicial complexes to illustrate these concepts for simplicity of presentation.

Given a simplicial complex $K$, a *discrete gradient vector* is a pair of simplices $(\sigma, \tau)$ such that $\sigma$ is a co-dimension one face of $\tau$; e.g., $\tau$ is an edge incident on a vertex $\sigma$, or $\tau$ is a triangle incident on an edge $\sigma$, and so on. Note that a discrete gradient vector is therefore a combinatorial pair, instead of a true vector. Nevertheless, a given pair, say $(\sigma, \tau)$, still indicates a "flow direction" from $\sigma$ to $\tau$, much like in the case of smooth Morse theory; see Fig. 3 where each discrete gradient vector $(\sigma, \tau)$ is indicated by a vector from $\sigma$ to $\tau$. A *V-path* is a sequence of simplices $\sigma_0, \tau_0, \sigma_1, \tau_1, \ldots, \sigma_k, \tau_k, \sigma_{k+1}$ such that for any $i \in [0, k]$, we have (i) the pair $(\sigma_i, \tau_i)$ is a discrete gradient vector, and (ii) $\sigma_{i+1}$ is a co-dimension one face of $\tau$. A V-path is *cyclic* if $a_0 = a_{k+1}$; otherwise, it is *acyclic*. See Fig. 3 for some examples.

Given a discrete gradient vector field $M$ of $K$, a simplex $\sigma$ is *critical* if it is not involved in any pair (ie., discrete gradient vector) in $M$; intuitively meaning that the gradient is "vanished" at $\sigma$. The index of this critical simplex is its dimension. Intuitively, given a triangulation $K$ of a 2D domain and a discrete gradient vector field $M$, critical vertices, critical edges, and critical triangles are analogous to minima (index-0 critical points), saddle (index-1) and maxima (index-2) for a smooth Morse function defined on this domain. The 1-unstable manifolds corresponding to saddles are V-paths connecting critical edges (saddle) to critical triangles (maxima). Each such V-path is a sequence of alternating edges and triangles. If $K$ is a triangulation of a $d$-dimensinal domain, such a V-path will be sequence of $(d-1)$-simplices and $d$-simplices, which can be expensive to compute and manipulate. Hence in practice, following [22, 24], we use the following trick: Instead of aiming to compute the 1-unstable manifold of an input function (viewed as a density field) $\rho$ to capture the mountain ridges, we calculate the *1-stable manifolds* of the negation of $\rho$, that is, for $-\rho$. The 1-stable manifolds capture "valley ridges" connecting the minima (bottom of the valleys) with saddles. In the discrete Morse setting, 1-stable manifolds correspond to V-paths connecting critical edges to critical vertices via sequence of alternating vertices and edges (regardless of the dimension of the domain), which are much easier to compute and maintain.

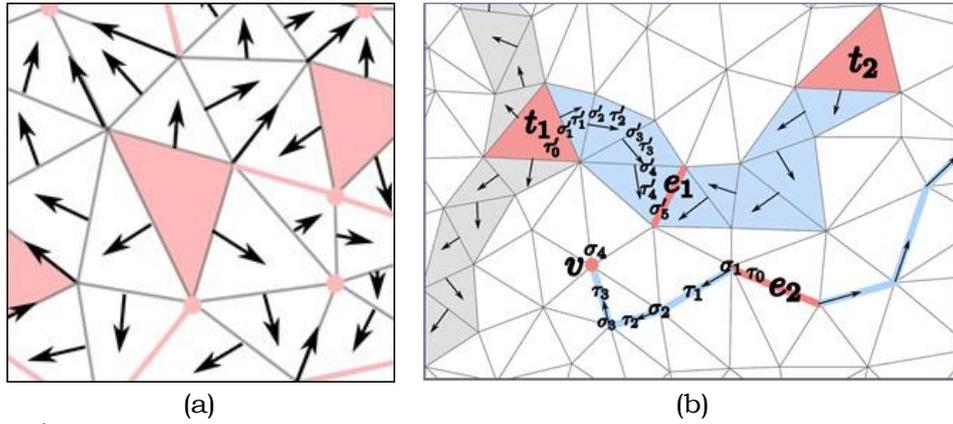

(a)                          (b)

**Supplementary Fig. 3 /** (a) A discrete gradient vector field. (b) Examples of V-paths analogous to 1-unstable manifolds (connecting $t_1/t_2$ to $e_1$) and 1-stable manifolds (connecting $v$ to $e_2$). (Image courtesy of [3].)

Finally, another advantage of using discrete Morse theory is that it provides a simple and combinatorial method for simplifying a given discrete gradient vector field $W$ to $W'$ to reduce the number of critical simplices. This is achieved via the so-called *Morse cancellation* operation: Specifically, a pair of critical simplices $\alpha$ and $\beta$ of dimension $p$ and $p+1$ (e.g, a critical vertex $\sigma$ and a critical edge $\beta$) is *cancellable* if there is a **unique** V-path connecting them. Given such a cancellable pair of critical pairs ($\alpha, \beta$), we can simply "invert" the direction of the gradient vectors in this path and this will produce a new gradient field where $\alpha$ and $\beta$ are no longer critical, while other critical simplices stay the same. This operation can be repeatedly performed as long as there are cancellable pairs of critical simplices to reduce the number of critical simplices, and thus simplify the discrete gradient vector field.

Note that any pair of critical simplices can be cancelled as long as the unique V-path condition is satisfied. To decide which pairs to simplify so as to remove the "noise" in the input density field, we will use another topological concept called the persistent homology which we briefly introduce next.

### S.1.3 Persistent homology

Persistent homology [3–5, 48–50] is a fundamental recent development in the field of computational topology, underlying many topological data analysis methods. Below we provide an intuitive description about it, to help explain how it can be used to measure importance of critical points. See [3] for more detailed introduction of these topics.

First we describe the idea for the smooth setting, where we assume that we are given a smooth function $f : X \to \mathbb{R}$ defined on a space $X$ (e.g, $X$ is a square for the case of 2D image). Now imagine we trace how the function $f$ evoles on $X$ via the following growing sequence of subspaces of $X$:

$$X_{t_1} \subseteq X_{t_2} \subseteq \cdots \subseteq X_{t_m} = X, \text{ with } t_1 < t_2 < \cdots t_m,$$

where $X_t := \{x \in X \mid f(x) \leq t\}$ is the so-called *sub-level set of $f$ at $t$*. This is called the *sub-level set filtration of $X$ w.r.t. $f$*, which intuitively sweeps $X$ by increasing $f$ values, and tracks the subspace $X_t$ already swept. In particular, during this process, sometimes *new topological features* (such as components, loops/handles, voids) can appear, and sometimes they disappear. These topological features can be captured algebraically by the so-called homology classes. The creation (birth) and destruction (death) of such features can be captured by the so-called *persistent homology* [49], which can be summarized by a simple summary, called the *persistence diagram*. More precisely, it turns out that the birth and death of topological features (homology classes) can only happen when the sublevel set $X_t$ sweeps through a critical point of $f$. We can therefore track the birth and death events by a collection of *persistent pairings* $\Pi_f = \{(v_b, v_d)\}$, where each pair $(v_b, v_d)$ contains the critical points where certain topological feature is created and killed. Their function values $f(v_b)$ and $f(v_d)$ are referred to as the *birth time* and *death time* of this feature. The corresponding collection of pairs of (birth-time, death-time) is called the *persistence diagram*, formally, $\text{dgm}(f) = \{(f(v_b), f(v_d)) \mid (v_b, v_d) \in \Pi_f\}$. Each *persistent point* $(f(v_b)), f(v_d))$, corresponding to the birth time and death time of some homological features through the filtration, gives rise to a point in the *birth-death plane*, as shown in Fig. 4 where we provide a simple example of 1D function. (For example, in this example, as we sweep pass minimum $x_3$, a new component is created in the sub-level set. This component is merged to an older component (created at $x_1$) when we sweep past critical point (maximum) $x_4$. This gives rise to a persistence pairing $(x_3, x_4)$ corresponding to the point $(f_3, f_4)$ in the persistence diagram.) The importance of the topological feature captured by the persistent pairing $(v_b, v_d)$ is captured by its *persistence*, defined as $|f(v_d) - f(v_d)|$, as it measures the "lifetime" of this topological feature through the filtration.

Equivalently, one can also represent a persistent point ($t_b$, $t_d$) as an interval, called a *persistent bar*, and the collection of persistent points then give rise to a collection of "bars", giving rise to a representation called *persistent barcode*. In Fig.2 of the main text, we used the persistence barcode representation to make the relation to the persistent pairing giving rise to each bar more clear.

Finally, we note that there is an algorithm [9] based on matrix reduction to compute persistent homology, persistent pairings and the resulting persistence diagram.

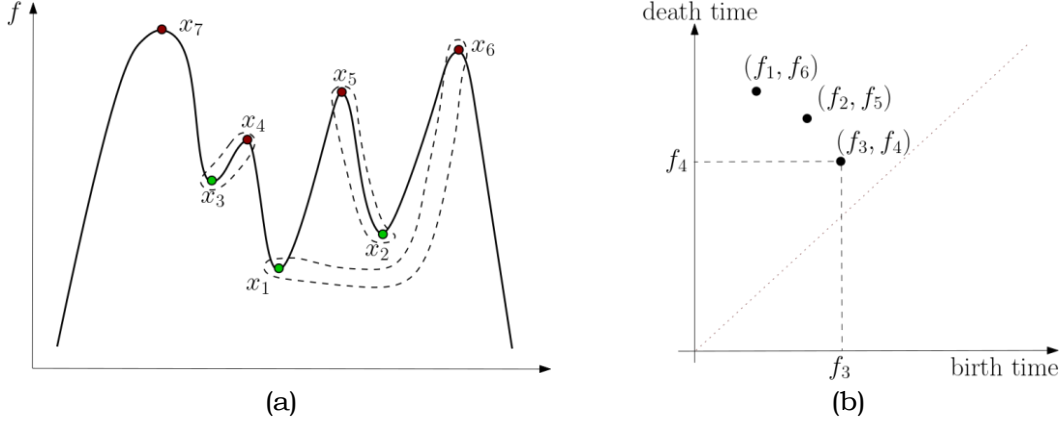

**Supplementary Fig. 4 /** (a) A simple 1D function $f : \mathrm{R} \to \mathrm{R}$. Its persistence pairings of critical points are marked by the dotted curves: $\Pi_f = \{(x_1, x_6), (x_2, x_5), (x_3, x_4), \ldots\}$. (b) shows its corresponding persistence diagram dgm($f$) = $\{(f_1, f_6), (f_2, f_5), (f_3, f_4), \ldots\}$, where $f_i = f(x_i)$ for each $i \in [1, 6]$.

**Persistence algorithm in discrete setting.** In the discrete setting, suppose we are given a triangulation $K$ and a function $f : V \to \mathrm{R}$ defined at vertices of $K$. We can simulate the above sub-level set filtration by the so-called *lower-star filtration*. Specifically, one can think that the function $f$ on $V$ is extended to a function on all simplices in $K$ by $f(\sigma) = \max_{v \in \sigma} f(v)$, for each simplex $\sigma \in K$ (e.g, for an edge $\sigma$, its function value $f(\sigma)$ equals the larger function value of its two vertices). One can then take the sublevel-set filtration of this simplex-wise valued function. Next, we can run the standard persistence algorithm [9] to this filtration, and the output is a collection of *pairs of simplices* $\Pi_f$. For each pair of simplices $(\sigma, \tau) \in \Pi_f$, it intuitively captures the birth and death of some homological feature in the sublevel sets. Analogous to the smooth setting, this pair of simplices gives rise to a persistent point ($f(\sigma)$, $f(\tau)$) in the persistence diagram, and its persistence $per(\sigma, \tau) = |f(\tau) - f(\sigma)|$ measures the lifetime (importance) of this feature. We will use these pairs and their importance as captured by the persistence to guide the simplification of discrete gradient vector field (and thus the resulting 1-(un)stable manifolds).

### S.1.4 Persistence-guided discrete Morse based graph reconstruction algorithm

We now put all pieces together and introduce the persistence-guided discrete Morse based graph reconstruction algorithm [24, 44], denoted by DiMorSC ().

On the high level, given a density function $\rho : X \to \mathrm{R}$ on a domain $X \subset \mathrm{R}^d$, note that we will consider the negation $f = -\rho$ of the density function, and aim to compute the 1-stable manifolds (instead of 1-unstable manifolds) to capture the valley ridgets (instead of mountain ridges). This is for the purpose to simplify the manipulation of discrete gradient vector field in the discrete Morse setting. Intuitively, we first compute the persistence pairings of all critical points of the function $-\rho$. We will then simplify the density function $\rho$ by "canceling" those pairs of critical points with low persistence, and only consider the 1-stable manifolds of remainder saddles with large persistence bigger than a given threshold $\delta$. The union of such 1-stable manifolds will capture important valley ridges of $-\rho$ (and thus mountain ridges of the density map $\rho$), and is the output reconstructed graph.

The above intuition can be translated to the discrete Morse setting, and we present the resulting algorithm in Algorithm 1, which is based on the simplified algorithm proposed by [22].

The original algorithm takes a triangulation K of the domain of interests and a density function $\rho$ given at the vertices of K as input. In our case, since our inputs are 2D images, instead of a triangulation, we take K to be the 2-skeleton (vertices, edges, and squares) of the 2D-cubical complex of the domain and use PHAT to compute persistence directly for such cubic complex. This does not change any part of the algorithm. Note that the algorithm will take as input a user-defined persistence threshold $\tau$; only 1-stable manifolds of saddles (critical edges) with persistence larger than $\tau$ will be computed and output.

**Algorithm 1** $G$ = DiMorSC (K, $\rho$, $\tau$)

1: Persistence Computation
   - Compute persistence pairings induced by lower-star filtration of K with respect to -$\rho$
2: Obtain Simplified Discrete Gradient Vector Field
   - Initialize trivial vector field
   - For each persistence pair, perform cancellation if possible and persistence $\leq \tau$
3: Collect Output
   - compute the 1-stable manifold of each critical edge with persistence > $\tau$
4: **return** union of 1-stable manifolds as reconstructed graph

*Step 1.*

Given a 2D-cubic complex K with a density function $\rho : V \to \mathbb{R}$ defined at vertices $V$ of K, we first perform the persistence algorithm to the lower-star filtration of $\rho' = -\rho$. The output is a collection $\Pi_\rho$ of pairs of cells in K (e.g, vertex-edge pairs or edge-square pairs). Note that as mentioned earlier, we use the negation of the density map $\rho' = -\rho$ in our algorithm so that it is easier later to compute the V-path between critical points (minima) and critical edges (index-1 saddles). In our implementation, we adapted DIPHA[40] to compute persistence because it is a distributed persistent homology algorithm which helps to reduce computation time.

*Step 2.*

The second step of the algorithm is to compute and simplify discrete gradient vector field. In theory, one should start with an initial trivial vector field $W$ where there is no discrete gradient vectors (that is, all simplices are critical at the beginning). One can then go through the persistence pairs from the output $\Pi_\rho$ of Step-1: any persistence pair $(\sigma, \tau)$ with persistence $\leq \tau$ is considered as noise, and one then performs the Morse Cancellation operation to cancel them if possible. In the end, this would give rise to a new discrete gradient vector field $W'$, where all low-persistence critical simplices are removed (if possible).

However, to implement this idea, [22] shows that in fact one does not need to explicitly perform Morse cancellations. Instead, all that is needed is to calculate the spanning forest that is made up of all negative edges (edges that are paired with vertices in $\Pi_f$) with persistence less than or equal to $\tau$. Positive edges (edges paired with a square) and edges with persistence greater than $\tau$ are not part of the spanning forest. No explicit discrete gradient vector field needs to be computed nor maintained. This spanning forest contains sufficient information for the Step 3 below. This step takes linear time once the persistence pairings are computed in Step 1.

*Step 3.*

The third step of the algorithm is to compute the 1-stable manifold of each critical edge whose persistence is at least $\tau$ in the simplified discrete gradient vector field. As shown in[22], for each such edge, the 1-stable manifold is equivalent to the union of the edge with the paths from both vertices to the sink of their corresponding tree in the spanning forest computed in Step 2. The union of all 1-stable manifolds is outputted by the algorithm as the reconstructed graph. Again we note that the 1-stable manifolds of -$\rho$ are analogous to the 1-unstable manifolds (mountain ridges) of the density field $\rho$.

We refer the output of this algorithm as the *Morse skeleton graph*, $G$. As shown in Fig. 1, this skeleton graph will then feed to the Simplification Step of the entire pipeline, to further remove noise, false branches and output the final tree/forest summary and produce vectorized objects for quantification.

# S.2 Point Spread Function

In order to estimate an areal line density from the detected fragments, it is necessary to estimate the effective thickness of the optical plane through the sample when acquiring the microscopic image. This thickness is necessarily nonzero due to the finite wavelength of light, and its extent may be estimated by characterizing the point spread function (PSF) of the microscopic imaging systems involved. For the fluorescent Whole Slide Imaging scans using the Nanozoomer HT2.0, we carried out an empirical determination of the PSF, using small fluorescent beads sufficiently smaller than the light wavelength to act as point sources, and also compared with the theoretical PSF. These two estimates sufficiently coincided for our purposes that we adopted the theoretical PSF to estimate the optical plane thickness as detailed below.

## S.2.1 Theoretical PSF

Various tools exist in the literature for estimating a theoretical PSF for wide field fluorescent microscopy. We used the Richards & Wolf (1959) model as implemented in an open-source package for PSF generation using the Fiji plug-in tool (http://bigwww.epfl.ch/algorithms/psfgenerator/). We used the following input parameters needed in this tool, corresponding to the imaging setup:

- Refractive index: (air) 1
- Wavelength: 660nm
- NA: 0.75
- XY resolution: 460nm
- Z: 130nm

The result theoretical PSF is symmetric to the z=0 plane, forming a cone shape with FWHM (Full Width Half Maximum) of approximately 1.5$\mu m$ in axial (z) direction, as seen in Supplementary Fig. 5.

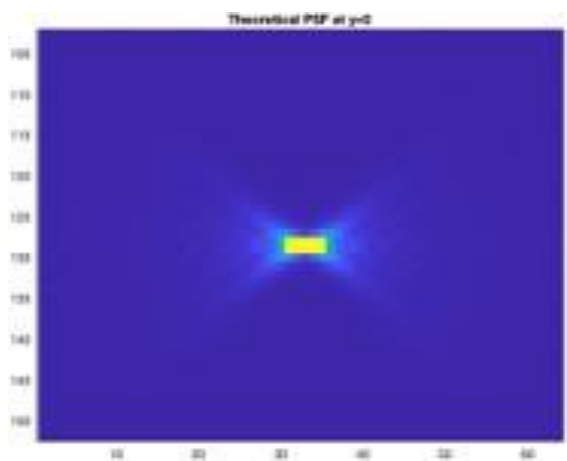

**Supplementary Fig. 5** / Theoretical PSF

## S.2.2 Experimental PSF

To obtain an empirical PSF of our imaging system, we imaged fluorescent nanobeads of 200nm. By aligning and averaging across 33 samples, we obtained an average experimental PSF as shown in Supplementary Fig. 6.

## S.2.3 Comparing the theoretical and experimental PSFs

We compared the theoretical and experimental PSFs by normalizing the intensity against the maximum point in the center, and overlaying the two functions within the same coordinate system. The experimental PSF can be seen to correspond approximately to the theoretical PSF in all planes of section as seen by comparing equivalent contour lines (Supplementary Fig. 7). We therefore adopted the FWHM of the theoretical PSF in the estimates in the text. Please note that this provides an overall multiplicative factor to the estimation of line density, so that errors in our estimation would only be reflected in an overall multiplicative factor, and not affect the relative distribution of line density across voxels in the image. For example, it would not affect the surprise estimate which proceeds from normalized probability densities. However, it would impact the estimation of the total length of axons emanating from the injection site. As discussed in the text, this is a biologically meaningful quantity, and therefore work the extra trouble incurred in trying to evaluate the PSF. The present work should be regarded as a first step in this direction, as there remains room for estimation of the corresponding stereological correction factors.

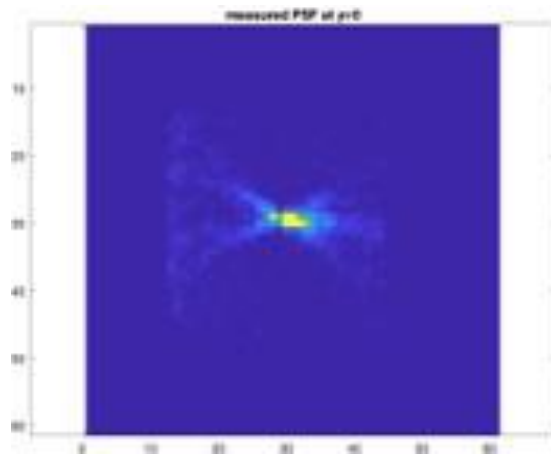

**Supplementary Fig. 6 /** Measured PSF

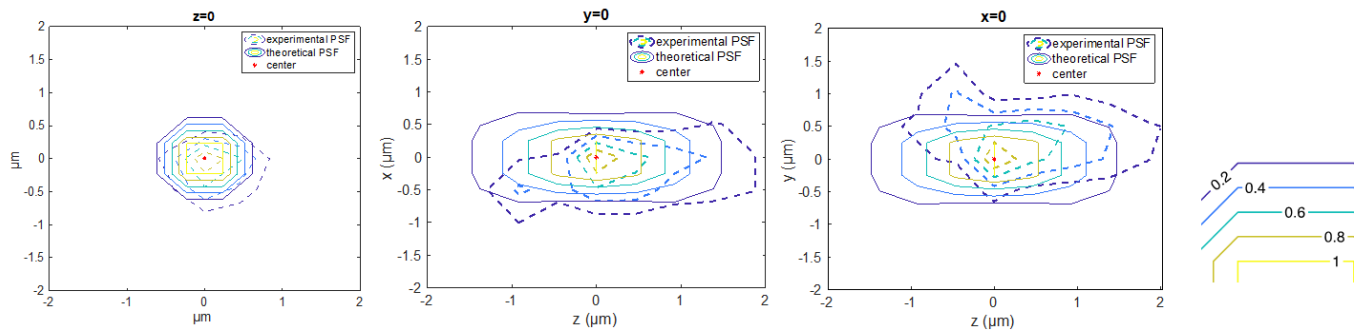

**Supplementary Fig. 7 /** Overlay of experimental and theoretical contours. The same color of the contour lines corresponds to the same level of normalised intensity level, taking the center as 1.